\documentclass{aa}  
\usepackage{graphicx}
\usepackage{txfonts}
\usepackage[allcolors=blue]{hyperref}
\begin{document} 

   \title{MAXI~J1348--630: Estimating the black hole mass and binary inclination using a scaling technique}

\author{ Lev Titarchuk \inst{1,2,3} 
\and
Elena Seifina\inst{3}
}

 \institute{$^1$ Dipartimento di Fisica, University di Ferrara, Via Saragat 1,
I-44122, Ferrara, Italy,  \email{titarchuk@fe.infn.it};\\
$^2$ Astro Space Center, Lebedev Physical Institute, Russian Academy of Science, 
Profsousnaya ul., 84/32, Moscow, 117997 Russia \\ 
$^3$ Lomonosov Moscow State University/Sternberg Astronomical Institute,
Universitetsky Prospect 13, Moscow, 119992, Russia,  \email{seif@sai.msu.ru}
} 

   \date{Received 
        ;       accepted 
}

  \abstract
{The multi-wavelength outburst activity in the recently discovered X-ray binary transient MAXI~J1348--630 has sparked a great deal of controversy about the characteristics of this binary and questions around whether the source contains a black hole (BH). Here,  we present the results of our analysis of the outburst of MAXI~J1348--630 using Swift/XRT data. We find that energy spectra in all spectral states can be modeled using a combination of Comptonization and Gaussian iron-line components.   In addition, we show that the X-ray photon index,  $\Gamma,$ is correlated with the mass accretion rate, $\dot M$. We find that $\Gamma$ increases monotonically  with $\dot M$ from the low-hard state to the high-soft state, and then becomes saturated at $\Gamma\sim$ 3. This index behavior is   similar to that exhibited by a number of other BH candidates. This  result represents observational evidence of  the presence of a BH in MAXI~J1348--630. We also show that the value of $\Gamma$ is correlated with   the quasi periodic oscillation frequency, $\nu_{L}$.   Based on this correlation, we applied a scaling method to estimate a BH mass of $14.8 \pm 0.9 M_{\odot}$, using the well-studied BH  binary XTE~J1550--564 as a reference source. The recent discovery of a giant dust scattering ring around MAXI~J1348--630 by SRG/eROSITA has refined  distance  estimates to this X-ray source. With this distance, we were able to estimate the disk inclination  $i = (65\pm 7)^{\circ}$ using the scaling technique for the correlation between $\Gamma$ and  normalization proportional to $\dot M$. We detected a specific behavior 
of the disk seed photon temperature, $kT_s$, immediately before the outburst: $kT_s$ initially decreases from 0.4 to 0.2 keV and increases only after the source transits to the outburst rise-maximum phase.  An initial decrease in $kT_s$ occurred simultaneously with an increase in the illumination fraction, $f$. We interpreted this effect in terms of the bulk motion Comptonization model. At the start of the outburst,   the Compton cloud (or "corona") is very extended and, thus, the seed photons injected to the corona from the relatively far-away disk region, where $kT_s$ is about 0.2--0.4 keV. While  $\dot M$ increases (or luminosity increases), the corona contracts, thus increasing the seed photon temperature, $kT_s$. It is possible that such a  decrease in $kT_s$  occurring simultaneously with an increase in the illumination fraction, $f,$ can be considered a signature of the readiness of a BH object to go into an outburst phase.
} 

\titlerunning{MAXI~J1348--630: a new BH mass and inclination estimates 
}

   \maketitle
%



\section{Introduction}

The X-ray transient MAXI~J1348--630 { was} discovered on January 26, 2019 with the Gas Slit Camera (GSC) of the Monitor of All-sky X-ray Image (MAXI) aboard the International Space Station 
\citep{Yatabe19} 
and  with the  Swift/BAT during its X-ray outburst. This 
outburst lasted for about four months. Further observations made it  possible to associate it with an activity of a black hole (BH) located at a distance of 3-4 kpc from the Earth \citep{Tominaga20}. 
The X-ray spectral and timing properties of this source estimates have been  studied in detail. Combined source spectra in the range of 1--150 keV from Swift/XRT, Swift/BAT, and MAXI/GSC were analyzed by \cite{Jana20} (hereafter, J20). They  appeared 
in the framework of a two-component advective-flow (TCAF) model  wherby the source evolved 
from the low-hard state (LHS) and through the intermediate state (IS) to the high-soft state 
[HSS, see, for example  \cite{mr06,ts09, st09}, 
hereafter ST09, for definitions of spectral states]. 
Spectral state evolution of MAXI~J1348--630 during its 
outburst is also confirmed by {\it NICER} observations [\citet{Zhang20}, hereafter Z20], 
which is similar to that previously observed for other BH transients.

Two 
 reflares of MAXI~J1348--630 were detected, occurring at the end of the main outburst and exhibiting peak fluxes that were one and two orders of magnitude lower than those of the main outburst, respectively, (Z20). These authors showed that the source remained in the hard state during 
 reflares, 
which were reminiscent of  so-called ``failed outbursts.'' 
Failed outbursts are usually less bright than regular outbursts at the peak and show no sign 
of transitions between spectral states~\citep{DelSanto16, Capitanio09,Sturner+Shrader05}. 
This phenomenon was also observed in many BH transients [\cite{Stiele+Kong16,Furst15,Sturner+Shrader05} and ST09]. 
The study of MAXI~J1348--630 with $NICER$ and $Swift$ also revealed different types of low-frequency QPOs 
at different phases of outburst (Z20 and J20). 
In fact, Z20 and J20 
also demonstrated it during the source power spectral evolution of MAXI~J1348--630, 
which was also similar to other known BH transients.  
 In Z20 and J20,  the results strongly supported a hypothesis that MAXI~J1348--630 contained a BH 
{and these authors estimated a BH mass as 8--12 M$_{\odot}$ (see Table~\ref{tab:previous_bh_mass}). 
}
 This estimate of a BH mass was made assuming the Shakura-Sunyaev accretion disk  (\cite{ss73}, hereafter SS73) and taking into account that the disk temperature, $T_d$. is inversely proportional to the fourth root of a BH mass. 
 Thus, Z20 estimated $T_d = 0.5-0.7$ keV during the outburst maximum  
 (see also \cite{Tominaga20}), which is slightly lower than that  ($T_d = 0.8-1.2$ keV) observed in other BHs  during the outburst peak \citep{Dunn11}.
 The lower disk temperature of MAXI~J1348--630 already indicates that it may harbor a BH with a mass { higher than 12 M$_{\odot}$ (Z20 and 
J20).} 

Recently, a X-ray spectroscopic analysis of MAXI~J1348--630 was performed by \cite{Zhang22} [hereafter, Z22] using Insight-HXMT and $Swift$ data. They found that MAXI~J1348--630 demonstrated the peculiar behavior during the ouburst rise in 2019. In Z22,  the source spectra were fitted  by the power-law plus the disk-blackbody components to find that in the soft intermediate and soft states, the object follows to the canonical relation of $L\sim T_d^4$ between the disk luminosity, $L,$ and the peak color temperature, $T_d$, at a constant inner radius, $R_{in}$. However, at other phases of the outburst, the behavior is more unusual and strongly deviates from the canonical evolution of known outbursts in BH transients. In particular, in the outburst rise phase, $R_{in}$ is smaller than in the soft state and the temperature, $T_d$, decreased to 0.5 keV. This contradicts  to the standard model of an optically thick inner disk that moves inward at the start of an outburst, becoming monotonically brighter and hotter and replacing the optically thin Comptonization region. Z22 associated this to a decrease in the hardening factor of disk emission with the outburst evolution at the rise phase.  In Z22, it was estimated that $T_d = 0.5-0.75$ keV during the outburst maximum. This low disk temperature also indicates that MAXI~J1348--630 may contain a BH with a mass above 12 M$_{\odot}$, which is consistent with the above conclusions on a BH mass by Z20, J20 and \cite{Tominaga20}.

To date, there are other estimates for the BH mass in MAXI~J1348--630  based on a large number of model parameters. In particular,  J20 (see section 2 and our Table~\ref{tab:previous_bh_mass}) applied five parameters for a BH mass estimates, such as  the Keplerian disk accretion rate ($\dot m$  in $\dot M_{\rm Ed}$ units), sub-Keplerian halo accretion rate ($\dot m_h$  in $\dot M_{\rm Ed}$ units), shock location ($X_{\rm s}$ in   units  of the Schwarzschild radius $r_{\rm s}=2GM_{BH} c^2$), and dimensionless shock compression ratio ($R = r_{+}/ r_{-}$, the  ratio of post-shock matter  density to the pre-shock matter density), which are essentially combined with  one instrument parameter, namely, the normalization constant ($N$).  \cite{Tominaga20} use the  {\tt Kerrbb} model (see our Table~\ref{tab:previous_bh_mass}), which gives relations among the spinning parameter, $a$, inclination angle, $i$, and a BH mass under a given source distance, mass accretion rate, and spectral hardening factor.  In Z20, the authors assumed the closer distance of 3 kpc (also supported by the large observed flux at the soft-to-hard state transition) and an intermediate inclination of $60^{\circ}$, for the average $R_{in}$  in the soft state using their model {\tt tbnew*(simpl*diskbb)}. 
With their parameters, namely, the disk temperature at the inner radius ($T_{in}$) in keV,  DISKBB normalization ($N_{diskbb}$),  photon  index, and scattered fraction (FracScatter), these authors obtained $R_{in} = 110 \pm 5$ km, which is consistent with the ISCO radius   for a non-spinning BH of 12 solar masses. Using the MAXI monitoring, \cite{Tominaga20} utilized  the measured evolution of the inner radius of the accretion disk  during the soft state to estimate a  BH mass (e.g.,  with  an improved distance), however, \cite{Lamer20} revised their inner disk radius measurements to  $R = 97\pm 13$ km , leading  an estimate of a BH mass as $11 \pm 2$  solar masses using the same model as in \citep{Tominaga20}. 

The BH mass estimates using these methods require an accurate knowledge of the distance to the source. 
However, the distance to MAXI~J1348--630 remains  highly debated.   Depending on the  method of measurement, the distance of MAXI~J1348--630 is 3--10 kpc [\cite{Nowak95, Maccarone03, VahdatMotlagh19,  Russell19,Tominaga20}  and J20). In Table~\ref{tab:previous_bh_mass}, we compare all  published values of a BH mass and the distance to MAXI~J1348--630. Recent observations taken during the first X-ray all-sky survey, using the SGR/eROSITA telescopes installed at the Spektr-RG space observatory, have led to a refinded distance of 
$D_{J1348}=3.4\pm0.4$ kpc \citep{Lamer20}.  This places  MAXI~J1348--630 in a region of a relatively low stellar density located between the spiral arms of Sagittarius and  Centauri of  the Milky Way. The new distance estimate has made it possible to now estimate a new BH mass as 11$\pm$2 M$_{\odot}$ (see Table~\ref{tab:previous_bh_mass}) based on the  assumption that its companion is a K-type star \citep{Lamer20}.

%
%
\begin{table}
 \caption{Previous BH mass estimates in MAXI~J1348--630}
 \label{tab:previous_bh_mass}
 \centering 
 \begin{tabular}{lcccl}
 \hline\hline                        
\\
  Method      &     M$_{BH},$     &     $i$,      &  $d$ \\
                   &     M$_{\odot}$  &     deg     &  kpc  \\
      \hline
 \\
TCAF model$^{a}$ & $9.1^{+1.6}_{-1.2}$  &  ...                &  5--10     \\
Kerrbb model$^{b}$ & 7                                       &  0                 &  4      \\
Luminosity$^{c}$   & 12                                       &  60                &  3  \\
Luminosity$^{d}$  & 11$\pm$2                            &  ...                &  3.4$\pm$0.4  \\
 \hline                                             
 \end{tabular}
 \tablebib{
(a) \cite{Jana20}; 
(b) \cite{Tominaga20}; 
(c) \cite{Zhang20}, and 
(d)  \cite{Lamer20}.}
 \end{table}


{ }


The existing estimates  for the  BH mass in this binary  (Table~\ref{tab:previous_bh_mass}) 
have a large scatter and are often based on a great number of model parameters, therefore 
these need to be refined. Since there is no dynamical estimate of the BH mass, a re-estimation based on models with a smaller number of parameters, as well as using alternative methods, is desirable.
%
In addition, it is important to prove the presence of a BH in this binary system -- not only on the basis of 
the mass value of the central object (> 3 M$_{\odot}$) but, for example,  using  other indicators as well,  such as  the detection of the constancy (or ``saturation'') of the photon index, $\Gamma,$ versus the  mass accretion rate, $\dot M,$  during X-ray outburst maximum, which is typical for other reliably established BHs [\cite{Stiele13} and ST09]. 
 In fact, the index-saturation effect when $\dot M$ 
is increased was already demonstrated in the early work by \cite{tz98}, offering a semi-analytical solution for the full kinetic equation; it was then was then  done using the Monte-Carlo simulations for BH sources 
(see \cite{LT99,LT11}.

%
%

\begin{figure*} 
 \centering
\includegraphics[width=14cm]{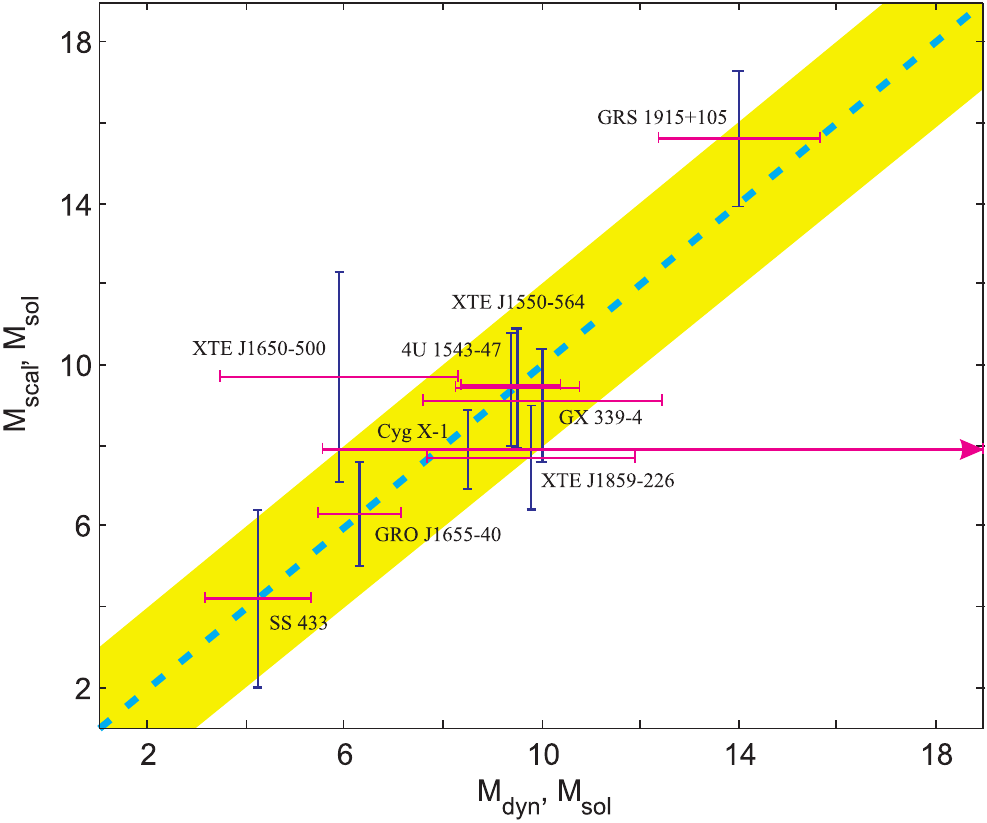}
\caption{BH mass from the scaling method (using saturation of the photon index with the mass accretion rate),
 $M_{scal}$, compared to the corresponding dynamical masses $M_{dyn}$ for stellar mass BHs. Data are taken from ST09, \citet{st03,st07,ST10} 
(for $M_{scal}$ values), 
and  
\cite{Greene01,Hjellming+Rupen95,Herrero95,Ninkov87,Munoz-Darias08,Hynes04,Park04,Orosz2002, SanchezFernandez99,Sobczak99,Petri08,Homan06,Filippenko+Chornock01,Miller-Jones21}
(for $M_{dyn}$ values). The yellow strip indicates a BH mass spread of 1.3~M$_{\odot}$.
The deviation from the yellow strip for Cyg~X--1 reflects new estimates of the BH mass (marked with an pink arrow), taking into account the strong wind in this binary and possibly associated with the refinement of the source distance (from 2.5 to 2.2 kpc) from the radio data -- due to which the BH mass estimate increased in Cyg~X--1 from 6.8--13.3$M_{\odot}$ to 21 $M_{\odot}$ \citep{Miller-Jones21}.
}
\label{Mdyn-Mscal}
 \end{figure*}

At this point,  we can estimate a  BH mass in MAXI~J1348--630 using the scaling technique. This technique was proposed by \cite{st07} (hereafter ST07) 
and developed by ST09. It was successfully tested in application to a large number of stellar mass BHs (e.g., \cite{STS14,tss10},  ST09,  see Fig.~\ref{Mdyn-Mscal}), as well as intermediate-mass BHs~\citep{TS16,TS17,STV17}, and supermassive BHs  \citep{STS16,sp09,ggt14,STU18,SCT18,TS21}. 

We note that according to ST07 and ST09, there are two   scaling methods:  based on the correlation between 
$\Gamma$ and  the quasi-periodic oscillation frequency  (QPO) $\nu_L$; and  based on the correlation between $\Gamma$ and  normalization of the spectrum proportional to $\dot M$ (see below Eqs. (\ref{bmc_norm})--(\ref{bmc_norm_lum}) for the definition of a mass accretion rate $\dot M$). For the first {method} ($\Gamma-\nu_L$), obtaining an estimate of the mass of the BH the distance to the source is not required (ST07); whereas for the second{ method} ($\Gamma-\dot M$), the source distance and  the { inclination} are needed  (ST09).

 For both methods, it is important for  the source to show a change in spectral states during the outburst  and a characteristic behavior of $\Gamma$; namely, a monotonic increase of $\Gamma$  { with} $\nu_L$ {or} $\dot M$ in the  LHS$\to$IS$\to$HSS transition and reaching a constant level  ({saturating}) at high values of  $\nu_L$ or $\dot M$. The saturation of $\Gamma$ (i.e., the so-called ``$\Gamma$-saturation phase'') during  the outburst is a specific signature  that supports the notion that this particular object containing a BH~\citep{tz98}. Indeed,  the $\Gamma$-saturation phase can be caused only by an accretion flow converging to the event horizon of a BH (see the Monte-Carlo simulation results in  \cite{LT99,LT11}).

The scaling method has advantages in determining a BH mass, namely:  using  X-ray data associated with the innermost regions of the accretion disk carry the direct information about a BH and  consequently, an   estimate  of a BH mass  is independent of  the distance to the object and its inclination 
[using  ($\Gamma-\nu_L$) correlation].

In this paper, based on {\it Swift} data analysis, we estimate a BH mass in MAXI~J1348--630 and the system's inclination, by applying the scaling technique. In \S 2 we provide details of our data analysis, while in \S \ref{sp_analysis} we present a description of the spectral models used to fit these data. In \S \ref{gamma-QPO_1550_scaling}--\ref{inclination1}  we focus on observational results and their interpretation. In  \S  4 
we  discuss  the  main results of the paper. In \S \ref{conclusions} we present our final conclusions.

%
%

\begin{figure*} 
 \centering
\includegraphics[width=14cm]{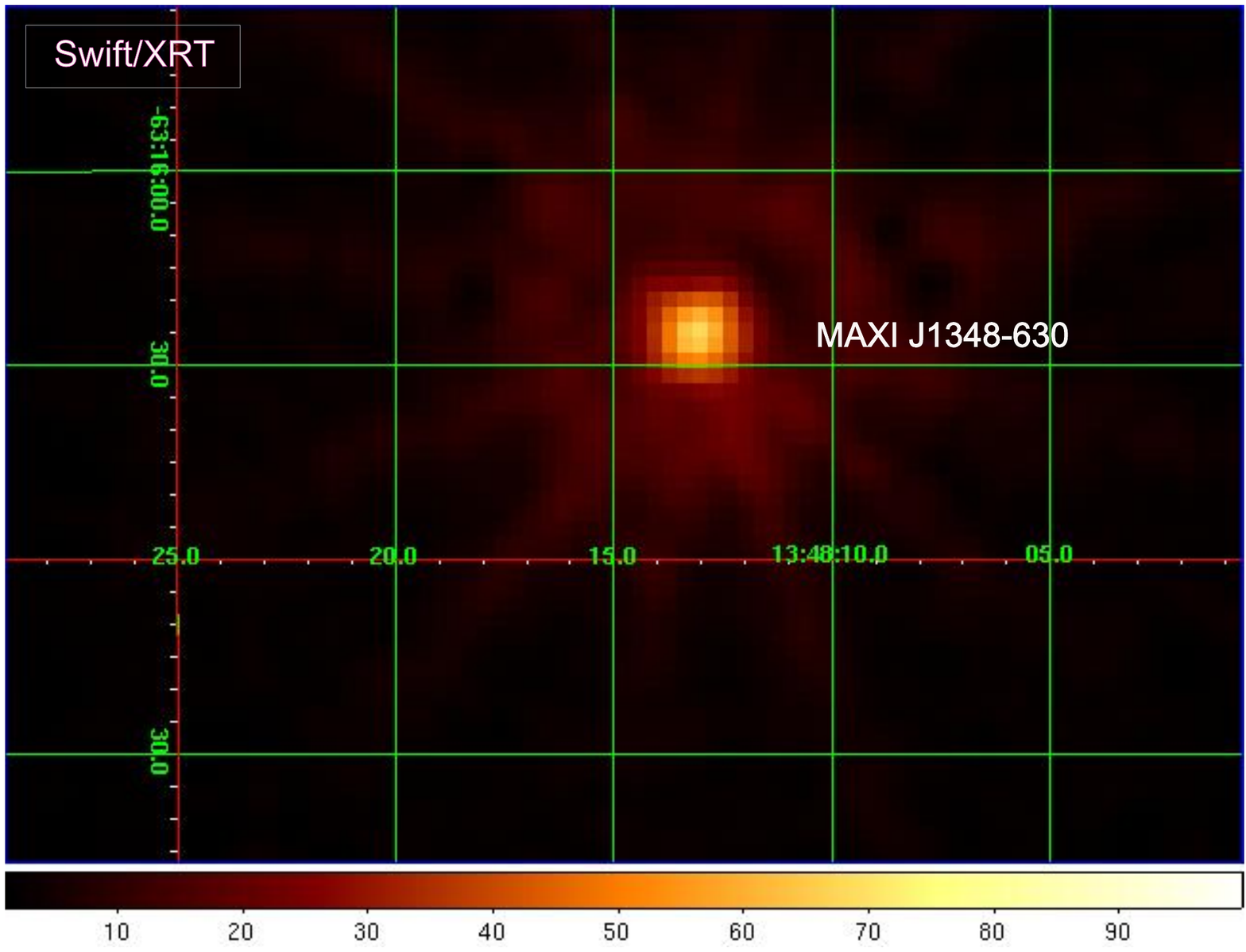}
\caption{Swift/XRT (0.3--10 keV) image of MAXI~J1348--30 accumulated from January 26, 2019 to October 17, 2020 with  an exposure of 52 ks. 
}
\label{image}
 \end{figure*}

%
%

\begin{figure*}
\centering
\includegraphics[width=14cm]{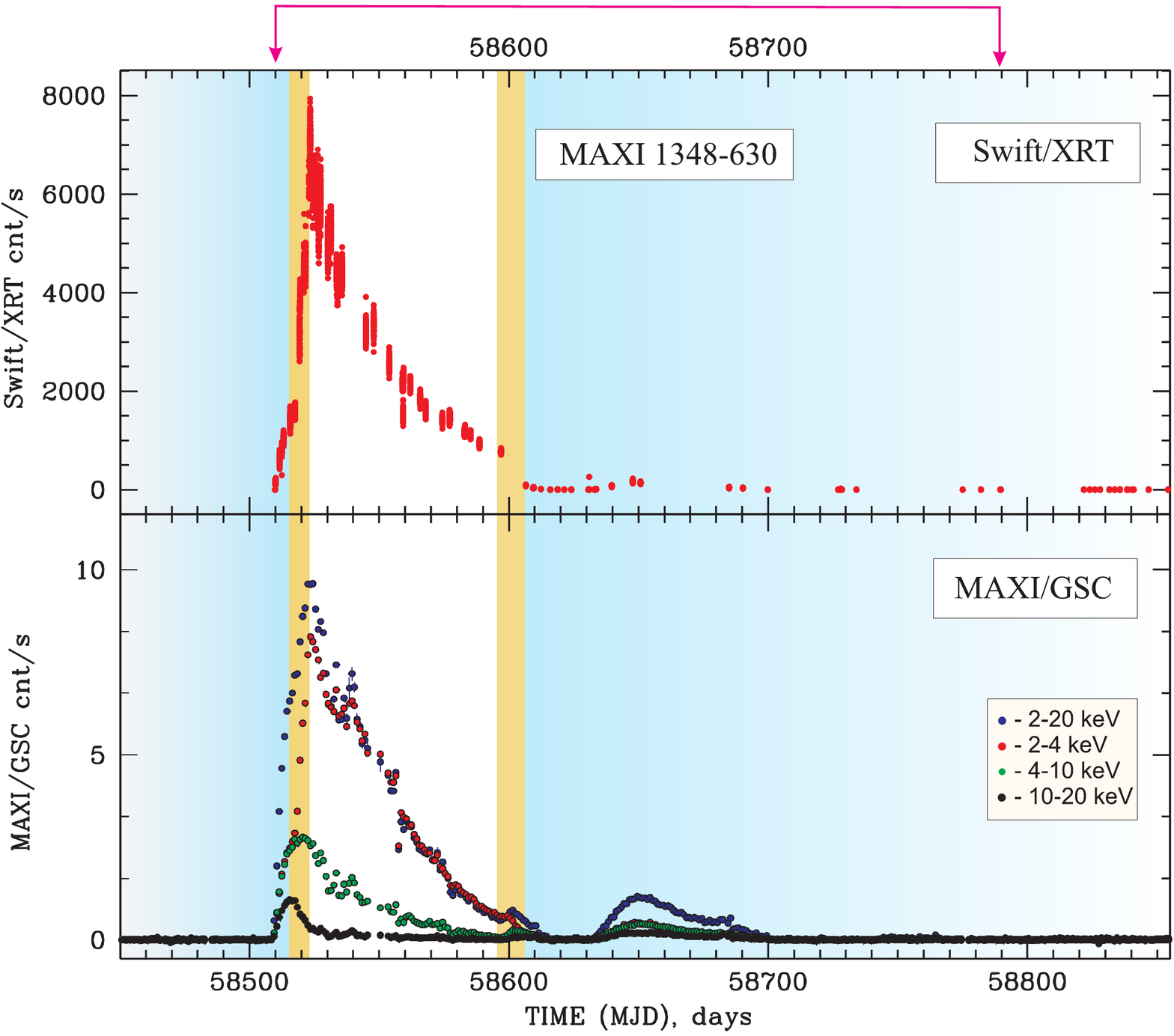}
\caption{Evolution of Swift/XRT (0.3--10 keV, {\it top panel}) and MAXI/GSC flux ({\it bottom panel}) in 2--20 keV, 2--4 keV, 4--10 keV and 10--20 keV energy ranges during 58450--58800 MJD observations of MAXI~J1348--630. Vertical bright blue, hazel, and white strips indicates  the LHS, IS and HSS spectral states, correspondingly  (for identification of spectral states see Sec.~\ref{sp_analysis}).  Pink arrow (at  top  of the panel) indicate the time interval under our study (from January to September,  2019) for  data sets listed in Table \ref{tab:list_Swift}.
 }
\label{MAXI_evol}
\end{figure*}

\section{ Data  Reduction  \label{data}}
\label{data}

{ Using {\it Swift}/XRT data in the 0.3--10 keV energy range,} we analyzed a total of 53 observations of the X-ray transient MAXI~J1348--630 during its outburst from January 26 to September 8, 2019.  
In Table~\ref{tab:list_Swift},  we report the log of observations for MAXI~J1348--630 used in our study. 

The data were processed using the HEASOFT v6.14, the tool {\tt xrtpipeline} v0.12.84 and the calibration files (CALDB version 4.1). The ancillary response files were created using {\tt xrtmkarf} v0.6.0 and exposure maps generated by {\tt xrtexpomap} v0.2.7.
Source events were accumulated within a circular region with radius of 20 pixels (1 pixel = 2.35 arcsec) centered at the position of MAXI~J1348--630 
[$\alpha$=13$^h$48$^m$12$^s$.7 and $\delta$=--63$^o16'26".8$, see  Kennea \& Negoro (2019)] 
{
when the X-ray source was faint (see  low-count plateau in   Fig. \ref{MAXI_evol}).
In the peak phase of  the outburst (MJD 58510 -- 58606), when the source became extremely bright and a {\tt  pile-up} effect occurred, we used an annulus extraction region with variable inner and outer radius (see column 11 in Table A1) to attenuate this problem \citep{Vaughan0}. 
}
We used XRT data both in the Windowed Timing (WT) mode during the bright part of the outburst (MJD 58509 -- 58720) and in the Photon Counting (PC) mode for the 
remaining observations when the X-ray source became sufficiently  faint ($\le$ 1 count/s). 
The background was estimated in a nearby source-free circular region. Using {\tt xselect} v2.4 task, the source and background light curves (0.01 s time resolution) and spectra were generated. 
The spectra were re-binned with 20 counts in each energy bin using the {\tt grppha} task in order to apply $\chi^2$-statistics. We also used the online XRT data product generator\footnote{http://www.swift.ac.uk/user\_objects/} to obtain the image of the source field of view in order to make a visual inspection and to
get rid of  a possible contamination from nearby sources    \citep{Evans07,Evans09}. In Figure~\ref{image}, we show an adaptively smoothed $Swift$/XRT (0.3 -- 10 keV) image  of the MAXI~J1348--630 field, obtained from 2019 January 26 to 2020 October 17, with  an exposure of 52 ks during MAXI~J1348--630, when it was was in outburst and quiescent states. 

%
%


\begin{table*}
 \caption{List of $Swift$ observations of MAXI~J1348--30 used in our analysis}              
 \label{tab:list_Swift}      
 \centering                                      
 \begin{tabular}{l l l l l c}          
 \hline\hline                        
  Obs. ID& Start time (UT)  & End time (UT) &MJD interval \\    
 \hline                                   
  00885807000$^{1,2,5}$       & 2019 Jan 26/12:07:07     & 2019 Jan 26 20:11:45   & 58509.47  -- 58509.84 &\\
  00885960000$^{1,5}$ & 2019 Jan 27 00:59:29    & 2019 Jan 27 00:59:35       & 58510.04 -- 58510.05 &\\
  00886266000$^{1,5,6}$       & 2019 Jan 28 15:05:37 & 2019 Jan 28 15:08:13 & 58511.58 -- 58511.63 &\\
  008864960000$^{1,5,6}$    & 2019 Jan 29 10:24:54    & 2019 Jan 29 10:25:01    & 58512.43 -- 58512.44 &\\
  00011107(001$^{1,5,6}$,002$^{3,4,5,6}$,003$^{3,6}$-023$^{5,6}$,024$^{4,5,6}$, & 2019 Jan 30        & 2019 Sep 8          & 58513.11 -- 58734.02 &\\
                          025-028,029$^{4,5,6}$,036$^{1,5}$, 037-055$^{5})$ &         &           &  &\\
  00088843(001-002)$^{1,5}$  & 2019 Feb 1       & 2019 Mar 8         & 58515.75 -- 58550.47 &\\
 \hline                                             
 \end{tabular}
 \tablebib{
(1) \cite{Jana20}; 
(2) \cite{Kennea_Negoro19}; 
(3) \cite{Bassi19}; 
(4) \cite{Tominaga20}; 
(5) \cite{Carotenuto21b} and   
(6) \cite{Zhang22}.
}
 \end{table*}

%
%
%
%
\begin{figure*}[ptbptbptb]
 \centering
\includegraphics[width=14cm]{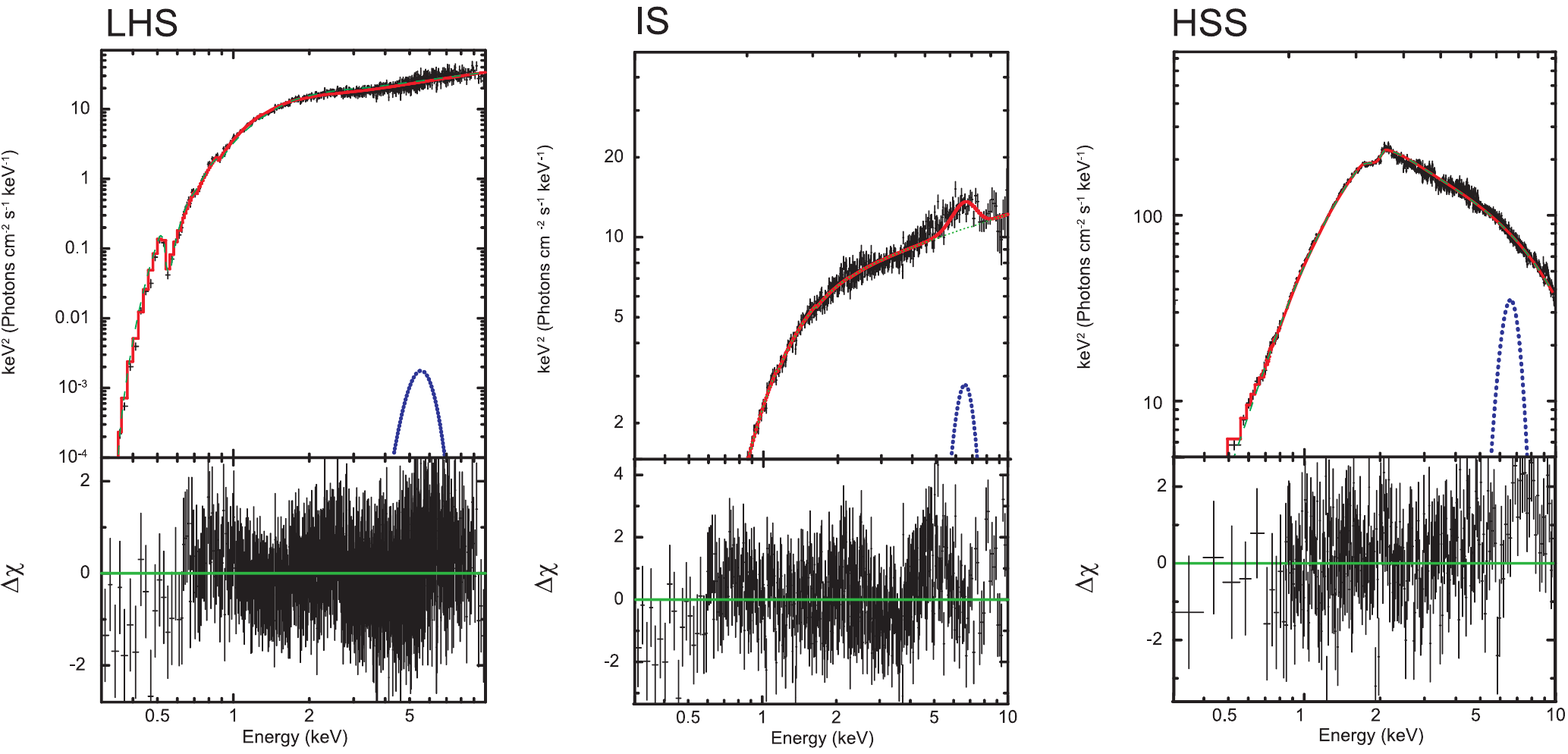}
\caption{
Evolution of MAXI~J1348--630 spectra,  fitted using 
the {\tt tbabs*(bmc+gauss)} model for three spectral states. Data are taken from the {\it Swift/XRT} observations: 00088843001 (MJD=58515.75, left), 00011107003 (MJD=58519, center), and 00011107007 (MJD=58524, right). Data are shown as black crosses and the spectral model components are displayed as dashed green and blue lines for the {\tt BMC} and {\tt gaussian}  components, respectively.  The resulting spectrum as a sum of these components are presented by red line.  
In the bottom panels, we show the corresponding $\Delta\chi$ versus photon energy (in keV). 
}
\label{sp_evol}
 \end{figure*}
 
The evolution of Swift/XRT in  0.3--10 keV  energy range during 58450--58800 MJD observations 
 is presented in the top panel of Fig.~\ref{MAXI_evol}. 
We also obtained 1-day bin MAXI light curves from the 
 2--20 keV band 
as well as the 2--4 keV,  4--10 keV and 10--20 keV bands
(Fig.~\ref{MAXI_evol}, bottom panel), through the MAXI ondemand Web interface\footnote{http://maxi.riken.jp/pubdata/v6m/J1348-632/index.html}. 

Only one outburst of this object is known to date (Fig.~\ref{MAXI_evol}), when it showed a rapid increase in X-ray luminosity (on the order of 10 days, during outburst rise phase) before reaching its peak luminosity, then followed by a slow luminosity decline over about four months (outburst decay phase). During this active phase, we can see   transitions between different spectral states (see color strip indications in Fig.~\ref{MAXI_evol}).



\section{Analysis and results \label{results}}

In this section, we present the results of spectral analysis during the 2019 outburst of MAXI~J1348--630 observed by $Swift$/XRT. 
In particular, we{ analyze}  how the X-ray spectrum { of the source} behaves, especially $\Gamma,$ during 
the {outburst} from MJD 58509 to 58734 MJD. 

%
%

\begin{figure*} 
 \centering
\includegraphics[width=16cm]{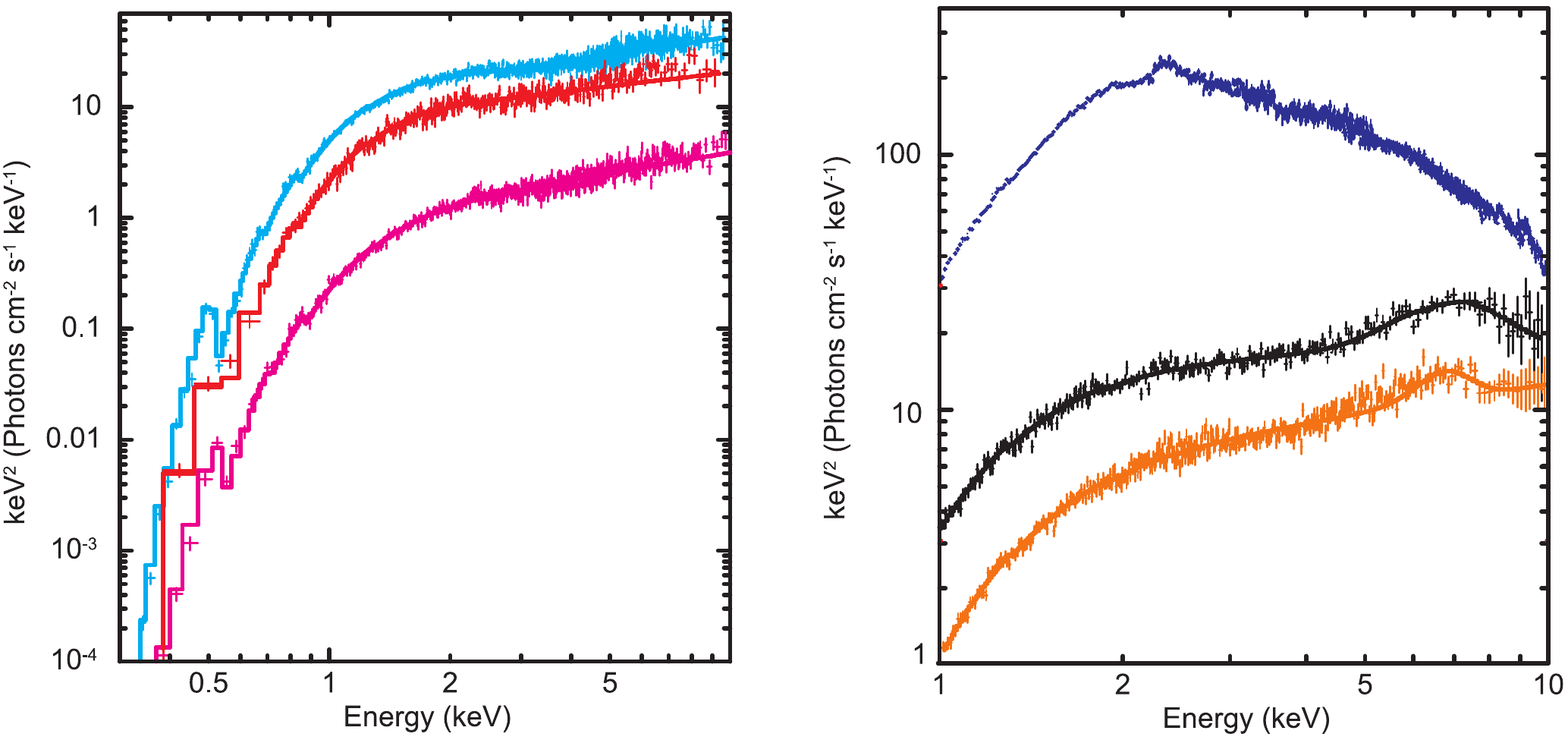}
\caption{Six representative $EF_E$ spectral diagrams during different spectral states of MAXI~J1348--30. Left panel: Data taken from {\it Swif/XRT} observations 0008884300 (bright blue, LHS), 0088496000 (red, LHS), and 0088596000 (pink, LHS). Right panel: Data taken  for 00011107003 (black, IS), 0088826000 (orange, IS), and 00011107007 (blue, HSS).}
\label{six_sp}
 \end{figure*}
 
\subsection{Spectral analysis \label{sp_analysis}}

To fit the energy spectra of this source, we used a {\tt XSPEC} model consisting of  the Comptonization (bulk-motion Comptonization, hereafter  BMC) component [see 
\cite{tz98, LT99}] 
and the iron line  ({\tt Gaussian}) components. We also used a multiplicative {\tt tbabs} model \citep{W00} ,
which takes into account  absorption by neutral material. 
We assume that  accretion onto a BH is described by two main zones [see, for example, Fig.~1 in \cite{TS21}]: a geometrically thin accretion disk, such as the standard Shakura-Sunyaev disk, (see SS73), 
and a transition layer  (TL), 
which is an intermediate link between the accretion disk, and a converging (bulk) region (see \cite{tf04}. The latter  
is assumed  to exist, at least below 3 Schwarzschild radii, $3R_S = 6GM_{\rm BH}/c^2$. 
The spectral model parameters are the equivalent hydrogen absorption 
column density, $N_H$; the photon index, $\Gamma$; whereas  $\log (A)$ is related to the Comptonized factor, $f$ [$={A}/{(1+A)}$], and the  color temperature and normalization of the seed photon blackbody component, $kT_s$ and   N$_{\rm bmc}$, respectively.  The Comptonized component  is additively combined with the {\tt gaussian} line model for which the fit parameters are the line energy, $E_{\rm line}$, and normalization, $N_{\rm line}$. The centroid of the {\tt gaussian} (Fe K$_{\alpha}$) line can vary from 6.3 to 6.9 keV; the Fe line width was varied from 0.1 to 1 keV and then subsequently fixed at 0.5 keV (as the best-fit value).  The parameter $\log(A)$ of the BMC component is fixed at 2 when the best-fit $\log(A)\gg 1$. In fact, for a sufficiently high $\log(A)\gg1$ (and, therefore, a high value A), the illumination factor $f = A/(1 + A)$ becomes a constant value close to 1 (that is, the same as in the case of $\log(A) = 2$).  $N_H$ was fixed at the level of {$0.64\times 10^{22}$ cm$^{-2}$  (Z20). 

Similarly to the ordinary bbody XSPEC model, the  BMC normalization  is a ratio of the source (disk) luminosity $L$ to the square of the distance, $d$  (ST09, see  Eq.~1 in that work): 

\begin{equation}
N_{bmc}=\biggl(\frac{L}{10^{39}\mathrm{erg/s}}\biggr)\biggl(\frac{10\,\mathrm{kpc}}{d}\biggr)^2.
\label{bmc_norm}
\end{equation}  

This encompasses  an important property of the BMC model. That is to say that using this model can lead to a correct
evaluatuation of the  normalization of the original ``seed'' component, which is presumably a correct $\dot M$ indicator~\citep{ST11}. In turn, we have:\  

\begin{equation}
L = \frac{GM_{BH}\dot M}{R_{*}}=\eta(r_{*})\dot m L_{Ed}.
\label{bmc_norm_lum}
\end{equation}  
Here $R_{*} = r_{*} R_S$ is an effective radius where the main energy release takes place in the disk, $R_S = 2GM/c^2$ is the Schwarzschild radius, $\eta = 1/(2r_{*})$, $\dot m = \dot M/\dot M_{crit}$ is the dimensionless $\dot M$ 
in units of the critical mass accretion rate, $\dot M_{crit} = L_{Ed}/c^2$, and $L_{Ed}$ is the Eddington luminosity. 
For the formulation of the  Comptonization problem,  we can look to \cite{tmk97,tz98,LT99,Borozdin99,st09}.  

The best-fit model parameters for  all states are shown  in Table~\ref{tab:fit_sp}}.  A systematic uncertainty of 1\% is intended to represent the instrumental flux calibration uncertainty and  has been applied to all analyzed $Swift$ spectra. 
 A spectral analysis of the $Swift$/XRT data{ fits} of MAXI~J1348--630, in principle, can provide a general picture of the spectral evolution. 
{ We can trace the change in the spectrum shape during the LHS-IS-HSS transition in} 
 Fig.~\ref{sp_evol}, which demonstrates three representative $E*F_E$ spectral diagrams 
for different states of MAXI~J1348--630. 
To identify the spectral states, we relied on the best-fit value of the photon index: LHS ($\Gamma<1.6$), IS ($1.6<\Gamma<1.8$) and HSS ($\Gamma>1.8$). 
The emergent spectra (see Fig.~\ref{sp_evol}) 
can be described  as   a sum of the low-energy blackbody  and its fraction convolved with the Comptonization Green function (CGF) [see 
 Eqs.~(16) and (B5) in \cite{st80}]. 
The HSS and IS spectra are characterized by a strong soft blackbody component (presumably related to the accretion  disk)  and a power law (as the hard tail of the CGF). 
In the LHS,  the Comptonization component is dominant and the  blackbody component is barely seen because the innermost part of the disk is fully covered by the  scattering media which has a 
Thomson optical depth  more than 2.

The general picture of the LHS-IS-HSS transition is illustrated in Figure~\ref{six_sp}, where we put together spectra of the LHS, IS, and HSS, to demonstrate the source spectral evolution from the high-soft to low-hard states based on the $Swift$ observations. Here, the data are presented in the left panel for LHS, taken from observations 0008884300 (bright blue), 0088496000 (red) and 0088596000 (pink); at the right panel for IS, we have 00011107003 (black), and 0088826000 (orange) as well as for HSS [00011107007 (blue). We should point out the fact that the HSS and IS spectra are characterized by a strong soft blackbody component and a power law extending up to 10 keV, while in the LHS spectrum, the Comptonization component is dominant and the blackbody component is barely seen. 

\begin{figure*}
 \centering
\includegraphics[width=17cm]{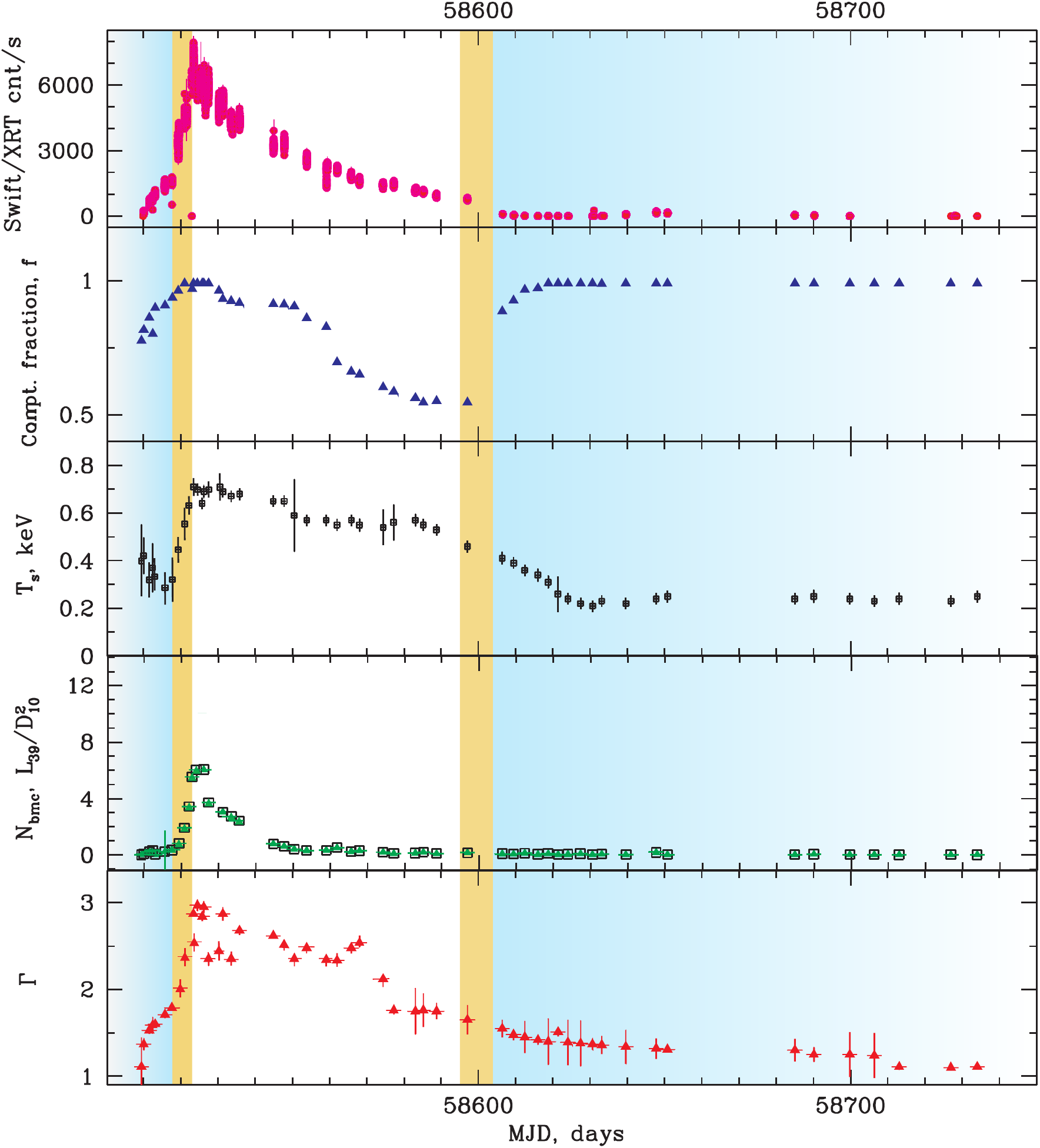}
\caption{
Temporal 
evolution of the Swift/XTE count rate (pink) from the LHS to the HSS and back, shown from top to bottom.  Comptonized fraction $f$ (blue), seed photon temperature $T_s$ (black), $N_{bmc}$ (green) and $\Gamma$ (red) during  the MJD=58509--58734 outburst transition set. We note that the Comptonized fraction $f=A/(A+1)$ characterizes the fraction of the soft photons illuminating the Compton cloud (CC), and subsequently up-scattered in the CC hot plasma. Vertical bright blue, hazel and white strips indicates  the LHS, IS, and HSS spectral states,  correspondingly (as in Fig.~\ref{MAXI_evol}). 
}
\label{frag}
 \end{figure*}

An analysis of  the $Swift$/XRT data fits (see  Figure~\ref{frag} and Table~\ref{tab:fit_sp}) showed that $\Gamma$  monotonically increases from 1.1 to 3 ($red$ points  in Fig. \ref{frag}), when the normalization of the BMC component (or $\dot M$) increases by a factor of 10 ($green$ { points, in Fig. ~\ref{frag}}) at the outburst rise phase (LHS-IS, MJD 58509-58512). At the rise-phase start of the outburst (LHS, MJD 58509 -- 58519), the spectra are gradually softened  with a increase in the total flux and $\Gamma$ { slowly} increases from 1.1 to 2, together with a slight decrease in the disk seed photon temperature, $kT_s$, from 0.45 to 0.2 keV ($black$  points, in Fig.~\ref{frag}). In this case, the contribution of the Comptonized component, $f,$ to the total flux increased from 0.7 to 0.9 ($blue$ { points, in Fig.~\ref{frag}}). Furthermore, at the end of the outburst-rise phase (HSS, MJD 58519 -- 58525), the spectra further softened, with an increase in the total flux, and $\Gamma$ continued to increase from 2 to 3, along with a slight increase in the disk seed photon temperature, $kT_s$, from 0.2 to 0.75 keV. At the same time, the contribution of the Comptonized component, $f,$ to the total flux remained almost constant, increasing from 0.9 to 1 ($blue$ { dots, in Fig.~\ref{frag}}). This interesting behavior 
of $kT_s$ and $f$ at the beginning of the outburst is further discussed in Sect.~\ref{discussion_maxi}. In turn, at the decay phase of the outburst (HSS, MJD 58525--58597), the spectra are gradually hardened,  with a decrease in the total flux, and $\Gamma$ { slowly} decreases from 
3. to 1.7, together with a slight decrease in the disk seed photon temperature $kT_s$ from 0.7 to 0.5 keV ($black$ { points, in Fig. ~\ref{frag}}). In this case, the contribution of the Comptonized component, $f,$ to the total flux decreased from 1 to 0.5 ($blue$ { points, in Fig. ~\ref{frag}}). When the source passed from the IS to  the LHS (MJD 58612), the temperature, $kT_s$,  dropped to $\sim$0.25 keV, and $\Gamma$  decreased to $\Gamma\sim$ 1.4, while the contribution of the Comptonized component to the total flux increased to $f$ = 1. After the source returned to the LHS (MJD 58606), the Comptonized component ($f$ = 1) prevailed in the spectra, while  $\Gamma$ changed  in the interval  1.1$-$1.4 and $kT_s$ was about 0.2 keV. 

%
%
\begin{table*}
 \caption{BH masses and distances}
 \label{tab:par_scal}
 \centering 
 \begin{tabular}{llllllc}
 \hline\hline                        
\\
Reference source   & $m_r^{(a)}$ (M$_{\odot})$ & $i_r^{(a)}$ (deg) & $d_r^{(b)}$ (kpc) &   References  \\
      \hline
\\
XTE~J1550--564  &   9.5$\pm$1.1 &  72$\pm$5    &  2.5           &   \cite{Orosz2002}; 
\cite{SanchezFernandez99} \\

 \hline\hline                        
\\
Target source   & $m_{t}^{(c)}$ (M$_{\odot}$) & $i_t^{(d)}$ (deg) & $d_t^{(b)}$ (kpc) &    \\
      \hline
\\
MAXI~J1348--630  & 14.8$\pm$0.9 &   65$\pm$5  & 3.4$\pm$ 0.4     & \cite{Lamer20} 
and this work \\
                            & 14.8$\pm$0.9 &   75$\pm$8  &  2.2$\pm$ 0.6   &  \cite{chauhan_21} 
and this work \\
\hline
\\
 \end{tabular}
 \tablefoot{
(a) Dynamically determined BH mass and system; (b) source distance found in literature; 
(c) scaled value found by $\Gamma-\nu_{L}$ correlation, and (d) scaled value found by the $\Gamma-N_{bmc}$ correlation of the present paper.
}
 \end{table*}
 
%
%
\begin{table*}
 \caption{Parameterizations of scaling patterns for reference and target sources}
 \label{tab:parametrization_scal}
 \centering 
 \begin{tabular}{lccccc||cccc}
 \hline\hline                        
\multicolumn{6}{c}{ $\Gamma$--$N_{bmc}$}&\multicolumn{4}{c}{ $\Gamma$-QPO}\\
  Reference source  &       $\cal A$ &     $\cal B$     &  $\cal D$  &    $N_{tr}$      & $\beta$  & A  & B               & $\nu_{tr}(Hz)$ & D \\
      \hline
\\
XTE~J1550-564 & 2.84$\pm$0.08 &  1.8$\pm$0.3    &  1.0 & 1.05$\pm$0.06   &   0.61$\pm$0.02  & 2.94$\pm$0.08& 1.27$\pm$0.02& 10.1$\pm$0.5 & 1.0\\

 \hline\hline                        
\\
  Target source     &      $\cal A$     &    $\cal B$   &  $\cal  D$  &   $N_{tr}$ & $\beta$ & A& B& $\nu_{tr}(Hz)$ & D\\
      \hline
\\
MAXI~J1348--630      & 2.96$\pm$0.07 & 0.52$\pm$0.02    & 1.0  &   1.2$\pm$0.1 & 0.50$\pm$0.03 & 2.93$\pm$0.09& 1.25$\pm$0.04& 6.48$\pm$0.07 & 1.0 \\

 \hline                                             
 \end{tabular}\\
 \tablefoot{
We use Eqs.~(3--6) for the analytical description of  $\Gamma$-QPO and  $\Gamma$--$N_{bmc}$ correlations, correspondingly.
}
 \end{table*}

\subsection{BH mass estimate}
\label{mass_estimate}

The previous BH mass estimates were made based on  luminosity determination or were based on the TSAF/Kerrbb  spectral model fitting [see \cite{Tominaga20,Russell19,Nowak95,Maccarone03,VahdatMotlagh19,Lamer20} and J20]. It is worth noting that a BH mass estimate using  these methods turns out to be highly dependent on the accuracy of the distance to the source. 

Now, we go on to estimate the BH mass ($M_{J1348}$) using scaling methods developed by ST09, based on $\Gamma-\nu_L$ (see our Sect.~\ref{gamma-QPO_1550_scaling}) and $\Gamma-N_{bmc}$ (Sect.~\ref{gamma-norm_1550_scaling}) correlations for the target source and the reference source. 
By using the $\Gamma-\nu_{L}$ correlation, we deal with  the method that offers the advantage of being independent of the binary inclination and source distance.

\subsubsection{Scaling of $\Gamma~vs~ \nu_{L}$ correlation for  MAXI J1348--630} 
 \label{gamma-QPO_1550_scaling}

\begin{figure*}
 \centering
 \includegraphics[width=14cm]{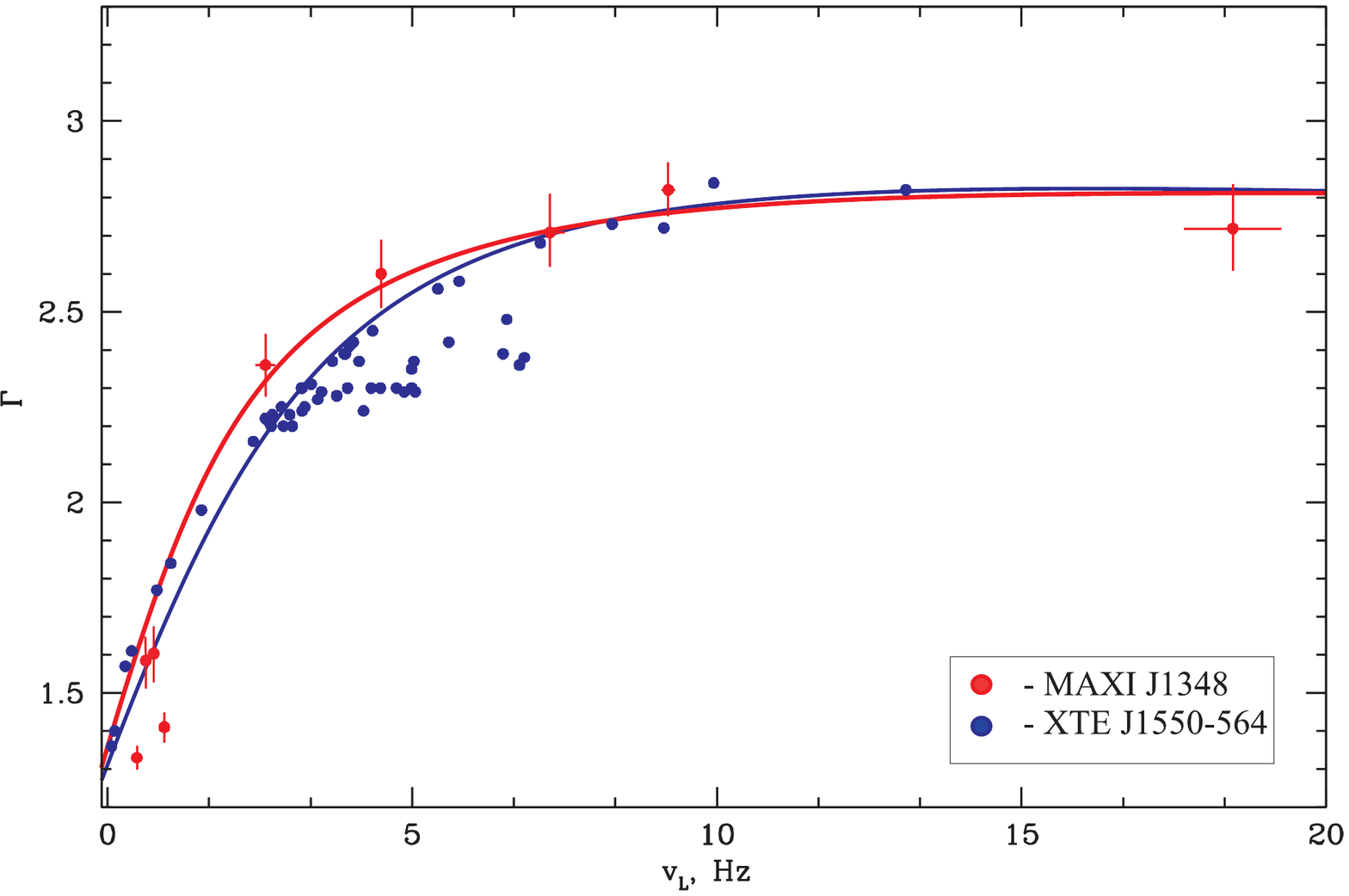}
      \caption{
{Photon index, $\Gamma,$ versus the QPO frequency, $\nu_{L}$, for  MAXI~J1348--630 (target source) 
and XTE~J1550--564  (reference source). We  use a specific function $f(\nu), $ detailed in Eq.~\ref{index-QPO_function} of Sect.~\ref{gamma-QPO_1550_scaling}, to fit the data, which gives us the corresponding solid lines.  We can see that  $\Gamma$ is correlated with  $\nu_{L}$ and  saturated for $\nu_{L}>6.8$ Hz  and $\nu_{L}>10$ Hz for  MAXI~J1348--630  and XTE~J1550--564, respectively. Data for XTE~J1550--564 were taken from ST09.}} 
\label{three_scal_2}
\end{figure*}

We obtained an estimate of a BH mass in MAXI~J1348--630 using the results of timing analysis by  Z20 (see their Table~1) in comparison with our results of spectral analysis for the close MJD dates (Table~A.1). 
We obtained  the  $\Gamma$ dependence on $\nu_{L}$ and demonstrated that  $\Gamma$ for  MAXI~J1348--630 is correlated with  $\nu_L$ and saturated at $\Gamma=$ 2.9$\pm$0.1 at frequencies above 6 Hz  (see Fig.~\ref{three_scal_2})\footnote{Hereafter, we consider the term  of the $\Gamma$ saturation as a constant  of  $\Gamma$ in the $\Gamma-\nu_{L}$ correlation track  at high values of $\nu_{L}$.}

In order to estimate $M_{J1348}$,  we chose Galactic source XTE~J1550--564 as the reference source (see a detailed explanation of the scaling  method in ST09). It is worth noting that XTE~J1550--564 is a source with the well-known parameters of the reference source: the spatial orientation (the inclination  $i_r$), BH mass ($m_r$), and  distance ($d_r$) were determined by a dynamic method (see Table~\ref{tab:par_scal} { and  \cite{Orosz2002}}).  Figure~\ref{three_scal_2} shows that $\Gamma-\nu_{L}$ correlations for MAXI J1348--630 and XTE J1550--564 follow a similar pattern  of $\Gamma$ with $\nu_{L}$: it  is linear and  monotonically increases  until it reaches a transition frequency, $\nu_{tr}$; whereas around  $\nu_{tr}$ the function $\Gamma-\nu_{L}$ smoothly transitions into a horizontal line.   It becomes a constant at a value $2.9\pm0.1,  $  which we assume  to be the $\Gamma$ saturation.

This behavior is well reproduced by the analytical function $f(\nu)$ (see ST09):
\begin{equation}
f(\nu)=A - ( D\cdot  B)\ln
\left(
\exp{\frac{\nu_{tr}-\nu} {D} + 1}
\right).
\label{index-QPO_function}
\end{equation}

Using this function, $f(\nu),$ we fit the observed $\Gamma-\nu_{L}$ correlations applying the least-squares method. Specifically, we minimized the sum of squared residuals of the function $f(\nu, A, B, D, \nu_{tr})$ from the observed $\Gamma-\nu_L$ track by selecting  values of the parameters $A, B, D, \nu_{tr}$ for which   this function, $f(\nu, A, B, D, \nu_{tr})$ is closest to  the observable  $\Gamma-\nu_L$ correlation.  

The significance of the parameters $A$ and $B$ follows from the asymptotic  of the function $f(\nu)$. For $\nu<<\nu_{tr}$  we see  that B is the slope of the correlation.  On the other hand,   for $\nu>>\nu_{tr} $ the function $f(\nu)$ equals to  $A$. From this it is immediately clear that the parameter $A$ is a value of the $\Gamma$-saturation level. We introduced the parameter $D$ in order to control how fast the transition occurs.

 Results after fitting the  $\Gamma-\nu_{L}$ correlations for these sources  using Eq. (\ref{index-QPO_function}), presented in Table~\ref{tab:parametrization_scal},  where $A =2.94\pm0.08$, $B=1.27\pm 0.01$, $\nu_{tr}=10.1\pm0.5$  Hz and $A =2.93\pm0.09$, $B=1.25\pm 0.04$, $\nu_{tr}=6.48\pm0.07$ Hz for XTE 1550-564 and MAXI~J1348--630, respectively. It is worth noting all these parameters are relatively close to each other. Numerous fit tests have shown that parameter $D$ varies slightly around 1 Hz, so we fixed it at 1.0 Hz for the convenience of the fitting procedure. We fit the correlations in order to get  these parameters,  $A$, $B$, $D,$ and  $\nu_{tr}$ (see Table~\ref{tab:parametrization_scal}), and thus  to obtain  the scaling factor, $s_{\nu}$ [Eqs.~(\ref{targetMass}) and (\ref{first_sc_law})]. Finally,  we estimated the BH mass in MAXI~J1348--630 applying the scaling method (see ST09,  Eq.~2 therein): 
 \begin{equation}
M_t=s_{\nu} \times M_r,
\label{targetMass}
\end{equation} 
where 

 \begin{equation}
s_\nu={\nu_{tr,r}}/{\nu_{tr,t}} 
\label{first_sc_law}
\end{equation} 
is the scaling factor and  subscripts $r$ and $t$ denote the reference and target sources.

For the reference source, we chose the data from the rise of the 1998 XTE~J1550--564 outburst, 
because its $\Gamma$-saturation level is approximately that of MAXI~J1348--630, $\Gamma_{sat}=2.9\pm0.1$. The best-fit curves are shown by  blue and red lines in Fig.~\ref{three_scal_2}. The inferred {uncertainty} (error bar) of $M_{J1348}$ is mostly affected by scattering and error bars of  the $\Gamma-\nu_{L}$ points.

As a result, the BH mass in MAXI J1348--630 is 
\begin{equation}
\begin{array}
{rcl}
M_{J1348}&=&s_{J1550\to J1348}*M_{J1550} = \\
&=&\frac{(\nu_{tr,J1550}/10.1 Hz)}{(\nu_{tr,J1348}/6.5 Hz)}(M_{J1550}/9.5 M_{\odot})=14.8\pm 0.9 M_{\odot},  
\end{array}
\label{BHmass}
\end{equation} 
\noindent using the scaling technique  and  the reference  BH mass, 9.5$\pm$ 1.1 $M_{\odot}$   (for XTE~J1550--564). For details, see Table~\ref{tab:par_scal}. 

\subsubsection{Scaling of $\Gamma-N_{bmc}$ correlation for  MAXI J1348--630} 
 \label{gamma-norm_1550_scaling}

The BH mass scaling method using the $\Gamma-N_{bmc}$ correlation is described in detail in  ST09. This method is aimed at: ({\it i}) searching for such a pair of BHs for which the $\Gamma$ correlates with {increasing} $N_{bmc}$ (which is proportional to $\dot M$, see ST09, Eqs.~4 and 7  therein)  and the saturation level $\Gamma_{sat}$ are the same and ({\it ii}) calculating the scaling coefficient $s_{N}$, which allows us to determine  a BH  mass of  the target object. It is worthwhile emphasizing that we needs  a ratio of  distances  for the target and reference sources in order  to estimate  a BH mass using  the following equation for the  scaling coefficient:   
\begin{equation}
 s_N=\frac{N_r}{N_t} =  \frac{m_r}{m_t} \frac{d_t^2}{d_r^2}{f_G}
\label{mass}
,\end{equation}
  where $N_r$, $N_t$ are normalizations of the spectra,   $m_t=M_t/M_{\odot}$,  $m_r=M_r/M_{\odot}$ are the dimensionless  BH masses with respect to solar, $d_t$ and $d_r$  are distances to  the target and reference sources, correspondingly.  We have a geometry factor of $f_G=\cos i_r/\cos i_t$, where $ i_r$ and $ i_r$ are the disk inclinations for   the reference  and target sources, respectively (see ST09, Eq.~7).  We compared the difference of the  disk  X-ray fluxes for the reference  and target sources,  in the direction towards the Earth observer. We used XTE~J1550--564 as the reference source. 

\begin{figure*}
 \centering
\includegraphics[width=14cm]{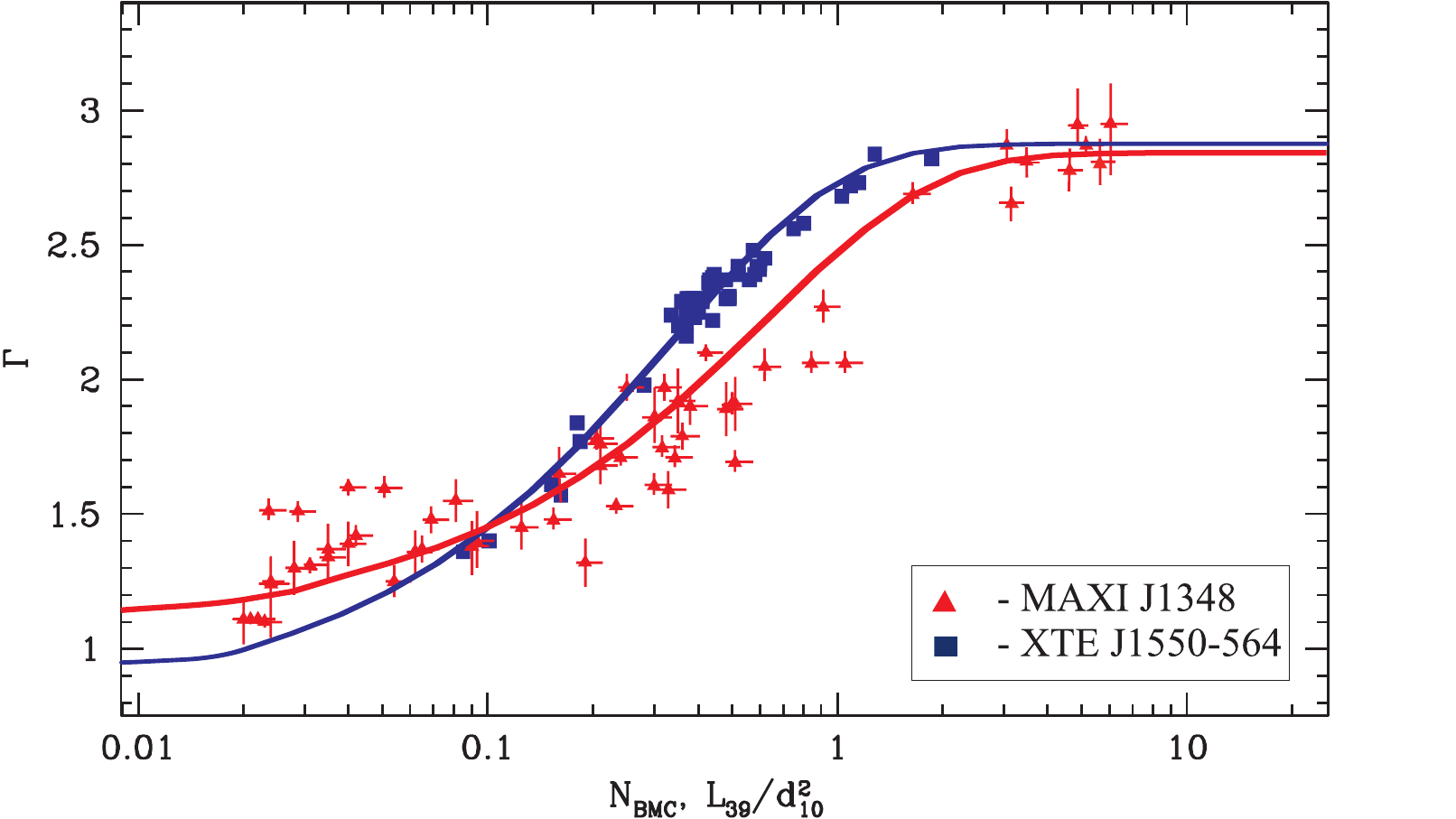}
      \caption{ $\Gamma$ vs  $N_{bmc}$ correlations those for MAXI~J1348--630 (target source) and for a BH  XTE~J1550--564 (reference source). We  use a specific function, $F(N),$ (see Eq.~8)  to fit the data, which gives us the corresponding  solid lines.
}
\label{three_scal_1}
\end{figure*}

In Fig. \ref{three_scal_1}, we demonstrate the $\Gamma$ versus  $N_{bmc}$, where  $N_{bmc}$  is presented in the units of $L_{39}/d^2_{10}$ ($L_{39}$ is the source luminosity in units of $10^{39}$ erg/s and $d_{10}$ is the distance to the source in units of 10 kpc). As  we can see,  the correlations of  both sources  are characterized by similar shapes and saturation levels, $\Gamma_{sat}\sim2.9\pm 0.2$.  In order to implement the scaling method, we used an analytical approximation $F(N)$ for the $\Gamma-N_{bmc}$ correlation, (see  ST09): 
\begin{equation}
F(N)= {\cal A} - ({\cal D}\cdot {\cal B})\ln\{\exp[(1.0 - (N/N_{tr})^{\beta})/{\cal D}] + 1\}, 
\label{scaling function_N}
\end{equation}
where $N=N_{bmc}$.
This function $F(N)$ is widely used for a description of the $\Gamma-N_{bmc}$ correlation [\cite{sp09}, ST09, \cite{ST10}, \cite{ggt14, STS14, STS16,STV17,STU18,SCT18}]. 
  
As a result of fitting the  observed correlation  by  this function $F(N),$ we obtained a set of the best-fit parameters $\cal A$, $\cal B$, $\cal D$, $N_{tr}$, and $\beta$ (see also  Table~\ref{tab:parametrization_scal}).   $\cal A$  is the $\Gamma$ saturation level for $\Gamma-N$ correlation and  $\cal B$ is the slope of this correlation. The parameter $\cal D$ controls how fast the transition occurs, $N_{tr}$ is the normalization at which the $\Gamma-N$ correlation levels of,f and $\beta$ is the power-law index of the part of the curve for lower argument values. 
 
%

We estimated a BH mass for MAXI~J1348--630 using Fig.~\ref{three_scal_2}. We  also fit   the correlation $\Gamma-N_{bmc}$ by Eq.~({\ref{scaling function_N}} ) to get  an estimate  for  $N^{J1348}_t = (1.2\pm 0.1)$ taken at the beginning of the $\Gamma$-saturation  part. We obtain $s_N=N_r/N_t$ applying $N$-values (see Table~\ref{tab:parametrization_scal}). Finally,  Eq.~(\ref{mass}) provides us a BH mass value with an accuracy of a factor $f_G$, namely,

\begin{equation}
m_t= f_G\frac{m_r}{s_N} \frac{d_t^2}{d_r^2}.
\label{mass_target1}
\end{equation}
We then obtain 
\begin{equation}
m_t= 20 f_G
\label{mass_target2}
\end{equation}
if we use values of $m_r=9.5$,  $d_t=3.39$ kpc and  $d_r=2.5$ kpc (Table 3) and  $c_N=(N_r/N_t)=0.875$ (Table 4).

Thus,  we estimate   $m_t=m_{J1348}$ with an accuracy of factor $f_G$ assuming $d_{J1348}$=3.39 kpc~\citep{Lamer20}. We used all the parameters of the formulas in Eqs. \ref{index-QPO_function} and \ref{scaling function_N} to fit  the observed  correlations $\Gamma$ versus $\nu_{QPO}$ and $\Gamma$ vs $N_{BMC}$  (see Table~\ref{tab:parametrization_scal}, Figs.~\ref{three_scal_2}-\ref{three_scal_1}).

 \subsection{Estimate of the orbital inclination for MAXI J1348--630  using the scaling technique \label{inclination1}}

 Now, we  estimate the inclination using a value of  the geometric factor $f_G$. We can evaluate $f_G$ more accurately, knowing  a value of a BH mass obtained using the $\Gamma-\nu_{L}$  scaling  [Eq.~(\ref{BHmass})]  and Eq.~(\ref{mass_target2}) that  for the $\Gamma-N$ scaling. Thus, equating a BH  mass using the $\Gamma-\nu_{L}$ correlation (Eq. 6)  to that for the  $\Gamma-N$  one (Eq. 10), we found that: 
\begin{equation}
f_G=0.74.
\label{f_G_value}
\end{equation}

 Moreover, we can estimate the inclination in MAXI~J1348--630 knowing the inclination $i$ of the XTE~J1550--54 (see Table~\ref{tab:par_scal}). Indeed, since:

\begin{equation}
f_G = (\cos i)_{J1550}/(\cos i)_{J1348}=0.74,
\label{fG3}
\end{equation}
\noindent we obtain the following:   
\begin{equation}
  i_{J1348} = (65 \pm 7)^{\circ}.
\label{inclination}
\end{equation}

In addition, we can evaluate a range of inclination values depending on the distance to MAXI~J1348--630; for example, for a distance range of 3--5 kpc, 
the inclination range can be estimated using the scaling method as then would be obtained using the Lamer et al. result to be  70--35$^{\circ}$. For the distance, 
 $d_t = 2.2^{+0.5}_{-0.6}$ kpc, obtained by \citep{chauhan_21}, taking into account the large uncertainties in the  source distance, we can only give a limiting range of the inclination as $75 ^{\circ}< i_t < 80^{\circ}$. This clearly  contradicts the absence of eclipses in the observed light curve (see Fig.~\ref{MAXI_evol}) in this binary system.

\section{Discussion \label{discussion_maxi}}

 The {Swift/XRT} data of the MAXI~J1348--630 { outburst} are well fitted by the BMC model for all analyzed LHS, IS and HSS spectra (see Figures~\ref{sp_evol} and \ref{frag}). Our results from the spectral analysis are consistent{ with previous results  by other} authors using various X-ray observations of MAXI~J1348--630 
 [see  Z20, J20 and  \cite{Lamer20}]. In particular, spectral parameter behavior {observed by us is} also consistent with those recently reported by \cite{Carotenuto21b}, who modeled the Swift/XRT spectra (1--10 keV) of MAXI~J1348--630 using the {\tt tbabs*powerlaw} model.  They found that the index $\Gamma$ changes from 1.5 to 2.4  during the outburst using their simplified model, {\tt tbabs*powerlaw}. 
However, these authors could not constrain $ \Gamma$ during the HSS and they fixed it at 2.4 (see their Figure 2). This outcome was not by chance, as  \cite{Carotenuto21b} used the phenomenological model {\tt tbabs*powerlaw} instead of the physical (Comptonization) model which we applied to the same data.

\cite{Tominaga20} modeled the MAXI~J1348--630 spectra detected by MAXI/GSC in 2--20 keV energy range using the {\tt tbabs*(simple*diskbb)} model and  found that $\Gamma$  changes from 1.5 to 2.9 -- this result is similar to ours (see our Figures \ref{three_scal_2}-\ref{three_scal_1}). Furthermore, \cite{Tominaga20} estimated the BH mass  by applying  the XSPEC kerrbb model and found that the values of $m_{BH} =M_{BH}/M_{\odot}$ were 7.0. It is worth pointing out that  the kerrbb model has five main parameters for the BH mass  estimate:\ the spinning parameter, $a$, inclination angle, $i$, and a black hole mass, $m_{BH}$, under a given source distance, mass accretion rate, and a spectral hardening factor. 
However, in our case, we used observational points of the $\Gamma$ versus QPO frequency correlation and only one the best-fit parameter, $m_{BH}$. Our scaling method allows us to  estimate $m_{J1348}$ as $14.8 \pm0.9,$ using  the XTE J1550  source as a reference.  

We compared our results with those obtained by Z20 based on NICER data. In Z20,  the MAXI~J1348--630 spectra were fitted during the outburst with the {\tt tbnew * (diskbb + nthcomp)} model and   then compared { with}  the {\tt tb-new*(simple*diskbb)} model results. In Z20,  higher values of $\Gamma$ were obtained throughout the whole outburst{ from MJD 58520 to 58635} ($\Gamma=1.5-3.8$), as compared to  our results ($\Gamma=1.1-3$). In particular, at the decay phase of the outburst (MJD 58525--58597), Z20 reported that  $\Gamma$ decreases only from 3.5 to 3, which is very different from  that found in this paper  (from 3 to 1.4). Moreover, a decrease of  $\Gamma$ in our estimate is { from 1.9 to 1.6 while   in Z20  it is from 3 to 1.7 {during MJD 58580--58610}. 
Z20  found a decrease in the temperature of the inner disk from $kT_{in} \sim$ 0.7 to 0.5 keV { (MJD 58520--58600)}, which is consistent with our results (from 0.7 to 0.4 keV). The decrease in the contribution of the Comptonized component, $f,$ to the total flux found by Z20 from  0.6 to 0.1 {(MJD 58520--58600)} is also consistent with the behavior of $\log(A)$ [$f=A/(1+A)$] (see Table~\ref{tab:fit_sp}) obtained in our spectral analysis as the X-ray flux decreases. However, {here} we found a wider range of  a decrease of $f$ from 1 to 0.1. {Again,  the contribution of the Comptonized component, $f,$ to the total flux increases up to 0.7 in {Z20, which is}}
somewhat more modest then  the result we obtained:\ ($f\sim 1$, {MJD 58600--58635}). 
 Furthermore, {Z20} found a sharp decrease in $kT_s$  from 0.5 to 0.2 keV when the source passed from the HSS to the LHS (MJD 58608), which is consistent with our results in general; however, we found a  decrease in $kT_s$ only from 0.4 to 0.3 keV. In { Z20,  the behavior of the spectral parameters is  similar with those found in our paper.  The general picture of the outburst in MAXI~J1348--630 is adequately reproduced in both models (compare our Fig. \ref{frag} to Fig. 5 in Z20). 

It is worth emphasizing that we have found a specific type of behavior with regard to the temperature of seed photons originating in the disk, $kT_s$,  at the very beginning of the outburst, just before the transition to the soft state (MJD 58510--58518). The temperature, $kT_s$, initially decreases (from 0.45 keV to 0.2 keV) and the Comptonized fraction, $f,$ increases from 0.75 to 0.9 during the initial rise in the low-hard state (Fig.~\ref{frag}). It is interesting to note that a similar behavior for $kT_s$ was found for the same source for these MJD dates by \cite{Zhang22}, although $kT_s$ decreased somewhat within different limits: from 0.75 to 0.5 keV. 
According to our spectral analysis the comptonized fraction $f$, at the same time, increases from 0.75 to 0.9 during the initial rise in LHS  (Fig. 6). We note that $f$ is a parameter associated with the relative size of the corona with a respect to a distance from the soft photon source (disk). As a result, its increase in $f$ immediately before the outburst (see panel 2 at the top of Fig. 6) simultaneously with the decrease in $kT_s$ points to an initial increase in the corona size.  Furthermore, at the end of the outburst-rise phase (HSS, MJD 58519 -- 58525), the spectra further soften with an increase in the total flux along with a slight increase in the disk seed photon temperature $kT_s$ from 0.2 to 0.75 keV. At the same time, the contribution of the Comptonized component, $f,$ to the total flux remained almost constant, increasing from 0.9 to 1 ($blue$ { dots, in Fig.~\ref{frag}}).  This effect has a simple physical interpretation in terms of the BMC model \citep{tz98, LT99}. At the beginning of the outburst (see Fig.~\ref{frag}), the  corona is very extended and, thus, seed photons are injected into the corona from a relatively far-away  region of the disk, where the disk temperature is about 0.2--0.4 keV. As the mass accretion rate increases (or the luminosity increases), the corona contracts and consequently  $kT_s$ increases (see an illustration of this phenomenon in Fig.~2 of \cite{STS14} and an analytical explanation of the evolution of the spectral index from the  low hard state to softer states in black holes in \cite{SCT18}. In this case, the parameter $f$ increases sharply (from 0.7 to 0.9). It is possible that such a decrease in  $kT_s$  and a simultaneous increase in $f$ immediately before the outburst can be considered a signature of the transition of a BH object into a flare (or of the readiness of the BH object to go into an outburst). This effect has also been observed in other black holes (e.g., GX~339--4, GRO~J1655--40, 4U~1543--47, XTE~J1550--564, XTE~J1650--500, H~1743--322, and XTE~J1859--226 [ST09], GRS~1915+105~\citep{ts09}, and 4U~1630--47~\citep{STS14} as well as in other studies of MAXI 1348 (Z22). Future observations of other BH sources that will demonstrate such behavior of $kT_s$ and $f$ on the eve of the flare may help verify the validity or even the universality of this signature.

Our results are also consistent, with the results of the spectral analysis by Z22 based on HXMT and $Swift$ data. However, those authors find, on average, a flatter  $\bar\Gamma\sim 2.4$, during outburst decay than our best-fit results ($\bar\Gamma\sim 2.7$). Perhaps the difference in $\Gamma$ compared to our results is due to the different energy range, or perhaps caused by a different spectral model. Throughout the outburst, their  $\Gamma$ ranges from 1.3 (ID=0214002002) to 2.9 (ID=0214002065, see their Table 1). This is in full agreement with our results ($\Gamma=1.1-2.9$). We note that when fitting the $Swift$ data, Z22 simply fixed $\Gamma$ at the canonical value during the outburst decay, $\Gamma = 2.5$, rather than looking for the best-fit solution (see their Table 2).

In Figures \ref{three_scal_2}-\ref{three_scal_1},  we show how smoothly $\Gamma$ evolves from the LHS to the HSS, with a clear indication of the saturation of $\Gamma$, that is, at least there is a sharp change in  the $\Gamma$ versus QPO frequency  slope  for high values  of $\nu_L$ and high mass accretion rates, $\dot M$,  respectively. 
This effect was predicted  semi-analytically  by \cite{tz98} and  then  using Monte-Carlo simulations by \cite{LT99,LT11}.
  Moreover it   was  confirmed  using the {\it RXTE} observations of Cyg X--1 (see  ST07 and ST09),  XTE~J1859--226, XTE~J1650--500, H1743--322, GX339--4, 
XTE~J1543--47 (ST09), 4U1630--47 \citep{STS14}, and GRS~1915+105 \citep{ts09}, as well as   observed in our study, which  allowed us to confirm the presence of a BH in MAXI~J1348--630. 

Using the scaling method ($\Gamma-\nu_{L}$) (ST07, ST09) we estimated the BH mass in  MAXI J1348--630  (see Fig.~\ref{three_scal_2} and Table~\ref{tab:par_scal}). The BH mass value 14.8$\pm$0.9 M$_{\odot}$ is relatively  far from  the previous BH mass estimate in MAXI J1348--630 of 7--12 M$_{\odot}$~\citep{Lamer20, Tominaga20}, J20 and Z20, see also our Table~\ref{tab:previous_bh_mass}  (compare Tables \ref{tab:previous_bh_mass} and \ref{tab:par_scal}). A possible reason for the discrepancy  between our  estimate  of a BH mass  and that based on luminosity  (Z20 and \cite{Lamer20})   is an  assumption by those authors  that the object emits at the Eddington luminosity regime, which, in fact may not be the case. As a result,  a BH mass value is naturally under-estimated. 

In contrast to Z20 and \cite{Lamer20}, we determined a BH mass using the first  scaling $\Gamma-\nu_L$ method and it was not necessary to know the distance to the source. Thus, the  main  advantages of the  scaling method ($\Gamma-\nu_L$)  to estimate a BH mass in comparison with other methods we used is that only one of the best-fit parameters is required, namely, a BH  mass, $M_{BH}$, for  at least  nine observational points of $\Gamma-\nu_L$ (see Fig \ref{three_scal_2}). It is worth noting that we also need a similar  $\Gamma-\nu_L$ correlation  with a known BH mass, which is  that of   XTE~J1550--564. 

We estimated the inclination, $i,$ by combining the two scaling methods ($\Gamma-\nu_{L}$ and $\Gamma-N_{bmc}$).  This estimate is  very different from the radio inclination estimate $i_{rad}<46^{\circ}$ \citep{Carotenuto21b}. A possible reason for this discrepancy may be related to the fact that the inclination estimate  from radio data is based on the difference in the brightness of the approaching jet and the receding jet based on the assumption that these jets are identical and oppositely directed,   which may not necessarily be the case \citep{chatterjee_20, Davis_Shane_Tchekhovskoy20}.  

Another reason for this discrepancy  using radio data and X-ray data   is their association  with  different geometric scales. It is known that the radio data is mostly formed in the outer part of the source, while the X-ray emission originates in the innermost region (TL).  Therefore, this estimate ($ i_{rad}<46^{\circ}$) may turn out to be unreliable for the inner part of the X-ray source. On the other hand, our scaling estimate [$ i = (65 \pm 5)^{\circ}$] can be a useful alternative  because we evaluate $i$ using X-ray emission  presumably originated  
in the TL.

\section{Conclusions}
\label{conclusions}

We studied the spectral evolution of MAXI~J1348--630 using fits  of the observed X-ray emission by $Swift$/XRT.  We demonstrated  that the energy spectra during all spectral states could be fitted by an additive model  consisting of the Comptonization (BMC) and Gaussian iron-line components. As a result, we found that $\Gamma$ monotonically increased  with $\nu_L$ and $\dot M$ during the transition from the LHS to the HSS and then became saturated at $\Gamma\sim$ 2.9 for high  $\nu_L$ and  $\dot M$ values. We applied these  correlations and found that   they were similar to those established in a number of other BH candidates and could be considered as an observational evidence for the presence of a BH in MAXI~J1348--630. 

We estimated a BH mass for MAXI~J1348--630 using the above scaling  methods (see   \S~3.2). In particular, the $\Gamma-\nu_{L}$ correlation scaling method, which relies on XTE~J1550--564 as a reference source, allowed us to estimate the  BH mass in MAXI~J1348--630  of  $M_{BH} = 14.8 \pm 0.9 $ M$_{\odot}$. It is important to emphasize once again   that the $\Gamma-\nu_{L}$ correlation is independent of the distance and inclination of the object and it is therefore are fairly accurate in the determination of a BH mass. 

We also detected a specific decrease in the disk seed photon temperature, $kT_s$, at the beginning of the outburs: $kT_s$ initially decreases from 0.4 to 0.2 keV and increases only after the source transits to the outburst rise-maximum phase. Initial decrease in $kT_s$ occurred simultaneously with an increase in the illumination fraction, $f$.  
We interpreted this effect in terms of the Comptonizaion model. Since the Compton cloud (or corona) is very extended at the outburst beginning and, thus, the seed photons injected to the corona from the relatively cold and peripheral disk region, where $kT_s$ is about 0.2--0.4 keV. While  $\dot M$ increases (or luminosity increases), the corona contracts and, thus, the seed photon temperature, $kT_s$, increases. It is possible that such a specific decrease in $kT_s$  taking place simultaneously with an increase in the illumination fraction, $f,$ can be considered to be a signature of the readiness of a BH object to go into an outburst.

~~~~~~
\section*{Acknowledgements}
We acknowledge support from UK $Swift$ Science Data Centre at the University of Leicester for supplied data. 
We thank the anonymous referee for the careful reading of the manuscript and for providing valuable comments.  We are very  happy to get a careful reading  and editing our  manuscript by Chris Shrader.  
This research has made using the  data and/or software provided by the High Energy Astrophysics Science Archive Research Center (HEASARC), which is a service of the Astrophysics Science Division at NASA/GSFC and the High Energy Astrophysics Division of the Smithsonian Astrophysical Observatory. The data used in this paper are public and available through the GSFC public archive at https://heasarc.gsfc.nasa.gov.
This work was made use of XRT and BAT data supplied by the UK $Swift$ Science Data Centre at
the University of Leicester\footnote{https://www.swift.ac.uk/swift\_ portal}, and MAXI data was provided by RIKEN, JAXA\footnote{http://maxi.riken.jp/mxondem}, and the MAXI team.





\begin{appendix}

\section{ Additional table \label{tab:fit_sp}}
%
%

\begin{table*}
 \caption{Best-fit parameters$^\dagger$ of the MAXI~J1348--630 spectra during 2019 outburst events 
observed by {\it Swift}.}
 \begin{tabular}{llccccccccccccc}
 \hline\hline  
Proposal ID     & MJD,  &  $\Gamma=\alpha+1$   &  $kT_{\rm s}$, &   $\log(A)^{\dagger\dagger}$ & $\nu_L^{\dagger\dagger\dagger}$, & $N_{\rm bmc}$, &  N$_{line}^{\dagger\dagger\dagger\dagger}$ & E$_{line}$, & $\chi^2_{\rm red}$(dof) & Ext.$^{\dagger\dagger\dagger\dagger\dagger}$ \\
00...                     &  day   &                                    &   keV               &                 & Hz &$L_{39}/d^2_{10}$&    &      keV   &        & region  \\
\hline  
885807000 &  58509.47 & 1.11$\pm$0.01 & 0.40(6)  & 0.54(8)  &  . . .& 0.02(1) & 0.10(2) &6.71(5)   &1.05 (710)  & 0--20 \\
885960000 &  58510.05 & 1.37$\pm$0.05 & 0.42(3)  & 1.62(2)  &  . . .& 0.06(1) & 0.14(5) &6.62(4)  & 0.97  (946)  & 3--30 \\
886266000 &  58511.58 & 1.53$\pm$0.03 & 0.32(3)  & 0.57(8)  &  . . .& 0.23(1) & 0.23(3) &6.74(7) & 0.91  (958) & 6--30 \\
886496000 &  58512.43 & 1.59$\pm$0.07 & 0.37(4)  & 0.60(1)    & 0.56(2) & 0.33(2) & 0.31(2) &6.58(6)  & 0.89  (881) &7--30 \\
011107001 &  58513.11 & 1.60$\pm$0.03 & 0.62(3)  & 0.64(2)  &  . . . & 0.04(1) & 0.29(6)  &6.90(4) &1.09  (898) & 10--30 \\
088843001 &  58515.75 & 1.71$\pm$0.03 & 0.29(8)  & 0.69(1)  &  . . .&  0.24(9) & 0.30(4) &6.68(5)  & 0.98  (958) & 12--30 \\
011107002 &  58517.67 & 1.79$\pm$0.03 & 0.31(9)  & 0.74(5)  &  . . .&  0.36(2) & 0.27(3) &6.83(7)  & 1.07  (846) &12--30 \\
011107003 &  58519.34 & 2.19$\pm$0.02 & 0.43(1)  & 0.82(2)  &  . . .&   0.48(2)  & 0.49(4) &6.64(5)  &1.06  (584)  &16-30 \\
011107004 &  58512.07 & 2.50$\pm$0.07 & 0.69(4)  & 0.95(1)  &  . . .&   0.38(1) & 0.52(8) &6.71(8)  & 1.00  (893) &31--50 \\
011107005 &  58523.00 & 2.79$\pm$0.04 & 0.70(3)  & 1.11(2)  &  . . . &  3.51(1) & 0.47(6) &6.85(4)  &  0.96  (894)   & 27--50 \\
011107006 &  58523.45 & 2.87$\pm$0.02 & 0.63(4)  & 2.00$^{\dagger\dagger}$&  . . .&   5.64(2)   & 0.61(7) &6.81(5)  &1.11   (552) & 32--50 \\
011107007 &  58524.53 & 2.97$\pm$0.05 & 0.70(2)  & 2.00$^{\dagger\dagger}$&  . . . &   6.27(2) & 0.84(5) &6.75(5)  & 1.05   (894) & 33--50 \\
011107008 &  58525.77 & 2.84$\pm$0.04 & 0.64(1)  & 2.00$^{\dagger\dagger}$&  . . .&   5.91(3)  & 0.71(5) &6.73(6)  & 0.74  (619)  & 32--50 \\
011107009 &  58526.25 & 2.85$\pm$0.05 & 0.69(5)  & 2.00$^{\dagger\dagger}$&  9.1(1)&  6.05(2)   & 0.83(9) &6.62(9)  &0.84   (525)  & 31--50 \\
011107010 &  58527.44 & 2.37$\pm$0.06 & 0.39(2)  & 2.00$^{\dagger\dagger}$&  2.6(2) &  4.90(2)  & 0.75(4) &6.40(5)  & 1.05   (894)  & 31--50 \\
011107011 &  58530.29 & 2.68$\pm$0.03 & 0.71(4)  & 1.42(7) &  4.49(1)&   0.21(2) & 4.52(8) &6.40(5)  &  1.02  (894) & 31--50 \\
011107012 &  58531.36 & 2.87$\pm$0.06 & 0.69(4)  & 1.14(5) &  . . .&   3.05(4) & 3.21(6) &6.42(7)  & 1.06   (486) &31--50 \\
011107013 &  58533.54 & 2.41$\pm$0.03 & 0.67(5)  & 1.09(6) &  . . .&   0.81(2) & 3.05(4) &6.56(3)  & 0.91   (451) &30--50 \\
011107014 &  58535.80 & 2.69$\pm$0.04 & 0.68(1)  & 1.04(1) &  . . .&   0.85(3) & 0.81(5) &6.64(8)  & 1.09   (440)  &30--50 \\
011107015 &  58544.91 & 2.68$\pm$0.02 & 0.69(4)  & 1.02(3) &  . . .&   0.79(2) & 0.74(3) &6.63(4)  & 1.12   (437)  &26--50 \\
011107016 &  58547.83 & 2.69$\pm$0.04 & 0.65(2)  & 1.01(2) &  7.3(2)&   0.62(1) & 0.76(4) &6.69(9)  & 1.08   (455)  &25--50 \\
088843002 &  58550.49 & 2.31$\pm$0.03 & 0.59(6)  & 0.98(7) &  . . .&   0.42(3)   & 0.80(6) &6.68(5)  &1.02  (476)  & 22--50 \\
011107017 &  58553.87 & 2.49$\pm$0.05 & 0.57(4)  & 0.79(6) &  . . .&   0.32(2)   & 0.75(5) &6.71(7)  &0.95   (426) & 17--30 \\
011107019 &  58559.13 & 2.41$\pm$0.02 & 0.57(1)  & 0.68(5) &  . . .&   0.35(2) & 0.71(4) &6.62(6)  & 1.07   (408) & 15--30 \\
011107020 &  58562.05 & 2.41$\pm$0.02 & 0.55(3)  & 0.36(2) &  . . .&   0.51(3)   & 0.72(3) &6.70(4)  & 1.11   (424) & 15--30 \\
011107021 &  58565.82 & 2.57$\pm$0.05 & 0.57(4)  & 0.29(4) &  . . .&   0.25(2)   & 0.69(6) &6.67(5)  & 1.09   (403) & 12--30 \\
011107022 &  58571.40 & 2.70$\pm$0.07 &  0.55(6) & 0.27(6) &  18.5(8)&   0.30(2)   & 0.63(3) &6.65(7)  & 1.05   (404) & 12--30 \\
011107024 &  58574.38 & 2.18$\pm$0.04 & 0.54(3)  & 0.18(2) &  . . .&   0.21(3)  & 0.67(4) &6.68(4)  & 0.96  (424)  & 11--30 \\
011107025 &  58577.24 & 1.76$\pm$0.03 & 0.56(3)  & 0.15(3) &  . . .&   0.11(1)   & 0.61(3) &6.61(5)  & 1.09  (403)  & 12--30 \\
011107026 &  58583.03 & 1.75$\pm$0.02 & 0.57(4)  & 0.11(6) &  . . .&   0.15(3) & 0.58(6) &6.59(7)  & 0.95   (386) & 11--30 \\
011107027 &  58585.21 & 1.76$\pm$0.05 & 0.55(1)  & 0.08(4) &  . . .&   0.21(2)   & 0.54(5) &6.63(4)  & 0.81   (364) & 11--30 \\
011107028 &  58588.73 & 1.75$\pm$0.07 & 0.53(2)  & 0.09(3) &  . . .&   0.10(3)   & 0.35(4) &6.49(5)  & 1.08   (347) & 12--30 \\
011107029 &  58597.04 & 1.6$\pm$0.1    & 0.46(5)  & 0.08(2) &  . . .&   0.16(2)  & 0.18(3) &6.48(7)  & 1.07   (316)  &10--30 \\
011107032 &  58606.39 & 1.55$\pm$0.08 & 0.41(3)  & 0.89(5) &  . . .&   0.08(3)   & 0.13(6) &6.42(8)  & 1.10   (394)  &3--30 \\
011107033 &  58609.47 & 1.48$\pm$0.05 & 0.29(1)  & 1.10(6) &  . . .&   0.07(2)   & 0.11(5) &6.58(5)   & 1.09   (437)  &0--20 \\
011107034 &  58612.44 & 1.45$\pm$0.09 & 0.25(5)  & 1.47(9) &  . . .&   0.13(4)   & 0.10(4) &6.57(7)  & 0.97   (258)  & 0--20 \\
011107035 &  58615.97 & 1.42$\pm$0.04 & 0.24(1)  & 1.54(6) &  . . .&   0.04(2)  & 0.14(3) &6.60(8)  & 1.06   (109)  & 0--20 \\
011107036 &  58618.74 & 1.40$\pm$0.08 & 0.21(6)  & 2.00$^{\dagger\dagger}$ &  . . .&  0.09(3)   & 0.12(6) &6.69(9)  &1.05 (72)  &0--20 \\
011107037 &  58621.39 & 1.51$\pm$0.04 & 0.26(3)  & 2.00$^{\dagger\dagger}$&  . . . &0.038(1) & 0.11(3) &6.57(7)  & 0.86  (50)  &0--20 \\
011107038 &  58624.07 & 1.39$\pm$0.09 & 0.24(2)  & 2.00$^{\dagger\dagger}$&  . . .&   0.04(1)   & 0.18(2) &6.55(7)  & 0.79   (45) &0--20 \\
011107039 &  58627.44 & 1.38$\pm$0.08 & 0.22(1)  & 2.00$^{\dagger\dagger}$&  . . .&  0.09(3)   & 0.13(5) &6.58(6)  & 0.81   (38) &0--20 \\
011107040 &  58630.69 & 1.37$\pm$0.05 & 0.21(4)  & 2.00$^{\dagger\dagger}$&  . . .&  0.03(1)   & 0.15(3) &6.61(9)  & 1.02   (33)  &0--20 \\
011107041 &  58633.15 & 1.36$\pm$0.08 & 0.23(3)  & 1.95(4) &  . . .&  0.06(2) & 0.19(2) &6.59(8)  & 1.09   (221) & 0--20 \\
011107042 &  58639.65 & 1.3$\pm$0.1    & 0.22(1)  & 2.00$^{\dagger\dagger}$&  . . .& 0.04(1)   & 0.11(3) &6.60(7)  & 1.02   (710)  &0--20 \\
011107043 &  58655.13 & 1.32$\pm$0.09 & 0.24(4)  & 2.00$^{\dagger\dagger}$&  0.9(2)&  0.19(4)  & 0.10(4) &6.62(9)  & 0.94   (470)  &0--20 \\
011107044 &  58655.62 & 1.31$\pm$0.03 & 0.25(6)  & 2.00$^{\dagger\dagger}$&  0.48(1)& 0.03(1)  & 0.18(1) &6.70(8)  & 0.89   (417) & 0--20 \\
011107045 &  58685.02 & 1.3$\pm$0.1    & 0.24(3)  & 2.00$^{\dagger\dagger}$&  . . .&  0.09(2)   & 0.10(3) &6.71(7)  & 1.06   (549) & 0--20 \\
011107047 &  58690.13 & 1.25$\pm$0.06 & 0.25(5)  & 2.00$^{\dagger\dagger}$&  . . .&  0.05(2)   & 0.13(2) &6.68(6)  & 1.05   (327) &0--20 \\
011107048 &  58699.83 & 1.2$\pm$0.1    & 0.24(1)  & 2.00$^{\dagger\dagger}$&  . . .&  0.02(1)  & 0.19(1) &6.63(7)  & 0.96   (59)  & 0--20 \\
011107049 &  58706.41 & 1.24$\pm$0.09 & 0.23(5)  & 2.00$^{\dagger\dagger}$&  . . .& 0.02(1) & 0.12(3) &6.65(7)  & 0.75   (43)  & 0--20 \\
011107050 &  58713.04 & 1.11$\pm$0.03 & 0.24(3)  & 2.00$^{\dagger\dagger}$&  . . .& 0.02(1) & 0.10(2) &6.60(6)  & 0.78   (32)  & 0--20 \\
011107052 &  58726.99 & 1.10$\pm$0.05 & 0.23(2)  & 2.00$^{\dagger\dagger}$&  . . .& 0.02(1) & 0.15(3) &6.64(9)  & 0.74   (30)  & 0--20 \\
011107055 &  58734.03 & 1.11$\pm$0.08 & 0.25(4)  & 2.00$^{\dagger\dagger}$&  . . .& 0.02(1) & 0.11(1) &6.62(4)  & 0.79   (30)  & 0--20 \\
\hline  
\end{tabular}
\tablefoot{ 
\\$^\dagger$ The spectral model is  {\tt tbabs*(bmc+gauss)};  
$^{\dagger\dagger}$ when the parameter $\log(A)>1$ then this parameter is fixed at a value of 2, because  for a sufficiently high $\log(A)$ (and, therefore, a high value A), the illumination factor $f = A/(1 + A)$ becomes a constant value close to 1;  
$^{\dagger\dagger\dagger}$ the centroid frequency taken from Z20; 
$^{\dagger\dagger\dagger\dagger}$  normalization parameter of the {Gaussian} model is in units of  {\rm total~photons~cm}$^{-2}$ {\rm s}$^{-1}$ in the line and $\sigma_{line}$ of {\it Gaussian} component is fixed to a value 0.5 keV (see comments in the text);   
$^{\dagger\dagger\dagger\dagger\dagger}$  the inner and outer radii of the annulus region for the spectrum extraction is indicated in pixels (1 pixel = 2.35 arcsec).
}
\end{table*}

\end{appendix}

\end{document}